\documentclass[lettersize,journal]{IEEEtran}
\usepackage{amsmath,amsfonts}
\usepackage{algorithmic}
\usepackage{algorithm}
\usepackage{array}
\usepackage{textcomp}
\usepackage{stfloats}
\usepackage{url}
\usepackage{verbatim}
\usepackage{graphicx}
\usepackage{subcaption}
\usepackage{cite}
\usepackage{multirow}
\usepackage{multicol}
\usepackage{booktabs}
\usepackage{color}
\usepackage[most]{tcolorbox}

\usepackage{mdframed}

\newmdenv[
  linecolor=red,
  linewidth=2pt,
  roundcorner=5pt,
  innertopmargin=5pt,
  innerbottommargin=5pt
]{redframe}

\usepackage[switch]{lineno} 

\ifdefined\isrevisionver
\newcommand{\modify}[2]{{\color{red}{#2}}}
\newcommand{\highlight}[1]{ \begin{tcolorbox}[colframe=red, colback=white, boxrule=1pt]
#1
\end{tcolorbox}}
\linenumbers
\else
\newcommand{\modify}[2]{#2}
\newcommand{\highlight}[1]{#1}
\fi

\newcommand{\donemodify}[2]{#2}
\newcommand{\donehighlight}[1]{#1}

\begin{document}

\title{TTSOps: A Closed-Loop Corpus Optimization Framework for Training Multi-Speaker TTS Models from Dark Data}

\author{Kentaro~Seki,~\IEEEmembership{Student Member,~IEEE}
        Shinnosuke~Takamichi,~\IEEEmembership{Member,~IEEE}
        Takaaki~Saeki,~\IEEEmembership{Member,~IEEE}
        and~Hiroshi~Saruwatari,~\IEEEmembership{Member,~IEEE}
\thanks{Manuscript received DD MMMM YY; accepted DD MMMM YY. Date of publication DD MMMM YY; date of current version DD MMMM YY.}
\thanks{This work was supported by JSPS KAKENHI 22H03639, 24KJ0860 and Moonshot R\&D Grant Number JPMJPS2011. }
\thanks{This work involved human subjects or animals in its research. Approval of all ethical and experimental procedures and protocols was granted by the Ethics Committee of the Graduate School of Information Science and Technology, the University of Tokyo, Japan, and performed in line with the subjective evaluations after obtaining informed consent from all the participants in the experiment.}
\thanks{At the time this study was conducted, all authors were with the Graduate School of Information Science and Technology, The University of Tokyo, Bunkyo-ku, Tokyo 113-8656, Japan.
Shinnosuke Takamichi is currently with the Graduate School of Information Science and Technology, The University of Tokyo, and the Graduate School of Science and Technology, Keio University, 3-14-1 Hiyoshi, Kohoku-ku, Yokohama-shi, Kanagawa 223-8522, Japan.}
\thanks{Corresponding author: Kentaro Seki (seki-kentaro922@g.ecc.u-tokyo.ac.jp).}}

\markboth{Journal of \LaTeX\ Class Files,~Vol.~14, No.~8, August~2021}%
{Shell \MakeLowercase{\textit{et al.}}: A Sample Article Using IEEEtran.cls for IEEE Journals}

\IEEEpubid{0000--0000/00\$00.00~\copyright~2021 IEEE}

\maketitle

\begin{abstract}
This paper presents TTSOps, a fully automated closed-loop framework for constructing multi-speaker text-to-speech (TTS) systems from noisy, uncurated web-scale speech data, often referred to as ``dark data,'' such as online videos.
Conventional TTS training pipelines require well-curated corpora with high acoustic quality and accurate text–speech alignment, which severely limits scalability, speaker diversity, and real-world applicability.
While recent studies have proposed acoustic-quality-based data selection techniques, they often overlook two critical aspects: (1) the inherent robustness of modern TTS models to noise, and (2) the potential contribution of perceptually low-quality yet informative samples.
To address these issues, TTSOps introduces a data-centric training pipeline that integrates three core components:
(1) automated data collection from dark data sources,
(2) utterance-level dynamic selection of data cleansing methods based on training data quality, and
(3) evaluation-in-the-loop data selection using automatically predicted mean opinion scores (MOS) to estimate each utterance’s impact on model performance.
Furthermore, TTSOps jointly optimizes the corpus and the TTS model in a closed-loop framework by dynamically adapting both data selection and data cleansing processes to the characteristics of the target TTS model.
Extensive experiments on Japanese YouTube data demonstrate that TTSOps outperforms conventional acoustic-quality-based baselines in both the naturalness and speaker diversity of synthesized speech.

\end{abstract}


\section{Introduction}
\IEEEPARstart{N}{eural} text-to-speech (TTS) systems have made remarkable advances in recent years, enabling the synthesis of speech with human-level naturalness and intelligibility~\cite{shen2018natural, popov2021grad, kim2021conditional, ju2024naturalspeech}.
These advancements are largely driven by deep generative models and the availability of large-scale, high-quality corpora.
Multi-speaker TTS models have further enabled flexible speaker control ~\cite{ping2018deep, jia2018transfer, cooper2020zero, casanova2022yourtts}.
However, they still rely heavily on curated datasets with clean audio and accurate text–speech alignments.
This reliance significantly limits speaker diversity and scalability in real-world applications.

In contrast, vast amounts of speech data already exist on the internet in the form of user-generated content, such as YouTube videos.
This so-called ``dark data''~\cite{trajanov2018dark, schembera2020dark} offers rich diversity in speaker identities, speaking styles, and recording environments, making it a potentially valuable resource for building multi-speaker TTS systems.
However, these real-world recordings are typically noisy, weakly labeled, and often misaligned with their transcripts, rendering them unsuitable for direct use in TTS training.
As a result, effective \textit{data selection} and \textit{data cleansing} are essential for leveraging dark data in high-quality multi-speaker TTS synthesis.

Conventional data selection methods for TTS often rely on predetermined metrics, especially acoustic quality (e.g., signal-to-noise ratio), to identify and retain ``clean'' utterances~\cite{zen2019libritts, bakhturina2021hi, dataset2021hui, rousseau2012ted}.
While this approach helps eliminate overtly corrupted or distorted samples, it assumes that acoustic quality is a sufficient proxy for the sample’s effectiveness in training a TTS model. 
However, recent advances in neural TTS have demonstrated that models exhibit increasing robustness to acoustic imperfections~\cite{zhang2021denoispeech, saeki2022drspeech}, and that even perceptually low-quality samples can meaningfully contribute to model performance.
As a result, selection strategies based solely on predetermined, model-agnostic criteria often fail to capture what truly matters for model training, leading to an increasing discrepancy between perceived acoustic quality and the actual effectiveness of data for model training\modify{—a property we refer to as \textit{training data quality}.}{. We refer to this property as \textit{training data quality}. 
Formally, the training data quality of an utterance is defined as its estimated contribution 
to the overall performance of the trained TTS model.}

In addition to data selection, most existing pipelines apply a single, uniform data cleansing method—such as denoising or restoration—to all training data, regardless of the individual characteristics of each utterance~\cite{koizumi2023libritts, watanabe2023coco}.
Although such approaches aim to enhance overall audio quality, they fail to account for the inherent heterogeneity of dark data, where the optimal data cleansing strategy may vary significantly across samples.
Some utterances may benefit from noise suppression, others from speech restoration, and some may already be clean—potentially degraded by unnecessary data cleansing.
This mismatch underscores the need for \textit{utterance-level adaptive data cleansing}, in which the optimal method is dynamically selected to maximize each sample’s utility for model training.

To address these limitations, we propose \textit{TTSOps}, a fully\newpage automated training pipeline for multi-speaker TTS that integrates both data selection and data cleansing into a unified framework.
The core idea of TTSOps is to assess the training data quality of each utterance—defined as its expected contribution to downstream model performance—and to use this assessment to inform two key decisions: (1) whether to include the utterance in training, and (2) which data cleansing method, if any, should be applied.
By embedding these decisions within a evaluation‑in‑the‑loop, TTSOps facilitates a model-aware and performance-driven approach to corpus construction from dark data.

To implement training-data-quality-based selection, TTSOps adopts an evaluation-in-the-loop strategy. 
For each utterance, a TTS model is trained using that utterance as part of the training data, followed by speech synthesis using the resulting model. 
The perceptual quality of the generated speech is then evaluated using a pre-trained pseudo MOS predictor. 
The resulting pseudo MOS score serves as a proxy for the utterance’s contribution to model output quality, thereby quantifying its utility as training data. 
By applying this process to a pool of candidate utterances, TTSOps selectively constructs a training corpus that prioritizes samples with the highest expected impact on learning.

In addition to data selection, TTSOps incorporates data-wise data cleansing switching into the training-evaluation loop.
For each utterance, multiple data cleansing variants—such as denoising, restoration, or no processing—are generated.
Each variant is then evaluated using the same pseudo MOS-based procedure described earlier: a TTS model is trained using the variant, synthetic speech is generated, and its training data quality is estimated.
The version with the highest estimated training data quality is selected and included in the final training corpus.
This unified loop enables TTSOps to jointly optimize both data selection and data cleansing in a model-aware manner, dynamically adapting to the characteristics of each utterance and the target TTS model.

We validate the effectiveness of TTSOps through experiments on actual YouTube data—a representative source of dark data characterized by high variability and minimal curation.
The results demonstrate that TTSOps consistently outperforms conventional approaches, including acoustic-quality-based data selection and fixed data cleansing pipelines, when evaluated in terms of predicted training data quality.
In particular, TTSOps improves the pseudo MOS of synthesized speech, increases the number of high-quality speakers, and enhances speaker diversity—all aligned with the optimization target used within the training loop. 
These findings highlight the value of training data quality as a guiding signal for both data selection and data cleansing, and demonstrate the promise of data-centric corpus construction, especially in noisy and heterogeneous settings.

This work extends our previous conference paper~\cite{seki2023text}, which introduced an evaluation-in-the-loop framework for training data selection. The prior study focused solely on utterance selection based on pseudo MOS scores, assuming a fixed data cleansing pipeline and evaluating the method under limited conditions (single TTS model, single corpus size). In this journal extension, we substantially broaden the methodology and experimental scope by introducing utterance-level adaptive data cleansing switching, validating the framework across multiple TTS architectures and corpus sizes, and conducting large-scale experiments on real-world YouTube data. These extensions demonstrate the generalizability and robustness of TTSOps, and further underscore the central role of training data quality in data-centric TTS system design.

\section{Related Work}
\label{sec:related_work}

\subsection{Multi-speaker and Noise-Robust TTS}
\label{sec:related:tts}
Recent advances in neural TTS have enabled models to synthesize speech for multiple speakers by leveraging speaker embeddings such as $x$-vectors~\cite{snyder2018x}. Models such as Deep Voice 3~\cite{ping2018deep} and various zero-shot adaptation techniques~\cite{cooper2020zero, casanova2022yourtts, mehta2024matcha, chen2025neural} have demonstrated the ability to generalize to unseen speakers with limited data. In parallel, noise-robust training approaches have been proposed to improve synthesis quality under degraded conditions. These include methods based on frame-level noise modeling~\cite{zhang2021denoispeech}, fine-grained noise control~\cite{nikitaras2022fine}, and degradation-robust representation learning~\cite{saeki2022drspeech}.

However, conventional TTS works tend to focus exclusively on improving model architectures, treating the training dataset as a fixed, pre-defined resource. In such approaches, data curation and preprocessing are performed as offline steps prior to model training, typically in a model-agnostic manner—despite growing evidence that the quality, diversity, and relevance of training data have a substantial impact on downstream performance. In contrast, our work adopts a data-centric perspective that treats corpus construction not as a static prerequisite, but as a dynamic, model-aware process that co-evolves with training.

\subsection{Data Selection for TTS Corpus Construction}
The performance of TTS models is highly sensitive to the quality and diversity of the training data. Since collecting well-curated corpora is expensive and time-consuming, several automatic data selection methods have been proposed to construct TTS corpora from speech resources such as ASR datasets~\cite{zen2019libritts, bakhturina2021hi, dataset2021hui, rousseau2012ted}. These methods typically rely on predetermined metrics such as signal-to-noise ratio (SNR) or perceptual quality indicators like NISQA~\cite{mittag2021nisqa} to select ``clean'' utterances.

However, as described in Section~\ref{sec:related:tts}, recent studies have shown that modern TTS models are increasingly robust to noisy inputs, and that even low-quality samples can contribute positively to learning.
In contrast to approaches that depend on predefined, model-agnostic quality metrics, our method evaluates each utterance based on its estimated training data quality—defined as its expected contribution to downstream model performance.

\subsection{Data Cleansing for TTS from Noisy Speech}
Speech enhancement has long been studied as a means to improve the perceptual quality of degraded speech signals~\cite{benesty2006speech}.
Related areas include dereverberation~\cite{nakatani2010speech, han2014learning}, speech separation or extraction from mixtures containing non-speech components~\cite{hu2020dccrn, rouard2022hybrid}, and analysis-by-synthesis methods that aim to remove distortions caused by recording devices~\cite{saeki2023selfremaster}.
Recently, deep generative approaches—often referred to as generative speech enhancement or speech restoration—have attracted increasing attention for their ability to produce natural-sounding outputs in subjective evaluations~\cite{su2021hifi, liu2022voicefixer, koizumi2023miipher}.

Recently, several TTS studies have leveraged these methods as data cleansing techniques~\cite{koizumi2023libritts, watanabe2023coco}, applying them to noisy corpora prior to model training.
However, most existing pipelines adopt a single, fixed data cleansing method across the entire dataset. 
This static strategy is often suboptimal, especially as more diverse and specialized restoration methods continue to emerge.
Selecting one best method in advance for every utterance is not only impractical, but also unnecessary.

In contrast, our framework eliminates the need for manual selection by dynamically combining multiple data cleansing methods in a model-aware manner. This allows us to construct an optimal TTS corpus in a fully automated fashion, while effectively adapting to the characteristics of each utterance and model.

\subsection{Automatic Assessment of Synthetic Speech Quality}
Perceptual evaluation of synthetic speech quality has traditionally relied on subjective mean opinion score (MOS) tests, which are labor-intensive and difficult to scale.
To overcome these limitations, a growing body of work has proposed automatic MOS prediction models that estimate perceived naturalness directly from audio signals without requiring reference signals or human raters.
Early approaches, such as AutoMOS~\cite{patton2016automos}, Quality-Net~\cite{fu2018quality}, and MOSNet~\cite{lo2019mosnet}, utilized raw waveforms or spectral features as input to recurrent or convolutional neural networks. 
Recent advances leverage self-supervised learning (SSL) representations, which are learned from large-scale unlabeled speech corpora, to dramatically enhance generalization and robustness~\cite{tseng2021utilizing, cooper2022generalization, saeki2022utmos}. 

The rapid progress in this field has been catalyzed by the annual VoiceMOS Challenge~\cite{huang2022voicemos, cooper2023voicemos, huang2024voicemos}.
Each edition introduces new tracks that test increasingly challenging generalization scenarios, such as out-of-domain (OOD) evaluation where models trained on English are used to assess Chinese speech~\cite{huang2022voicemos}, and zoomed-in comparisons where models must distinguish between highly natural-sounding synthetic speech samples~\cite{huang2024voicemos}.
These models are expected to continue evolving in both predictive accuracy and domain generalizability, facilitating broader applicability across languages, synthesis methods, and evaluation scenarios.

\subsection{Data-Centric AI and Its Application to Speech}
Data-centric AI emphasizes improving model performance by focusing on the quality, quantity and  diversity of training data, rather than solely optimizing model architectures~\cite{sambasivan2021everyone}.
In the field of speech technology, this perspective has motivated the creation of large-scale corpora from web-based sources, including GigaSpeech~\cite{chen2021gigaspeech}, JTubeSpeech~\cite{takamichi2021jtubespeech}, and J-CHAT~\cite{nakata2024j}.
However, these corpora are typically constructed using static data filtering pipelines, without incorporating model-aware evaluation or adaptive preprocessing.

The proposed TTSOps framework operationalizes data-centric AI principles in the context of TTS. By integrating data cleansing, data selection, and training within a closed-loop process guided by training data quality, TTSOps offers a scalable and task-adaptive approach to corpus construction from noisy, real-world speech data.

\begin{figure}[t]
\centering
\includegraphics[width=0.95\linewidth]{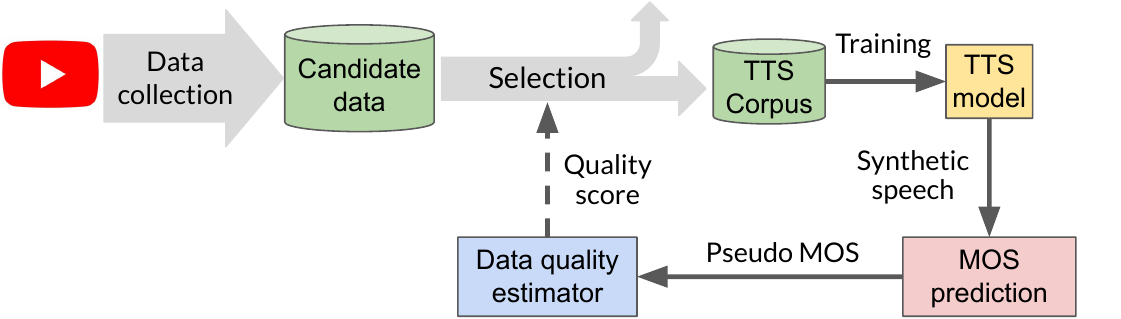}
\caption{
  Procedure of proposed data selection method based on training data quality.
  We evaluate training data quality of each utterance through loop of TTS model training and synthetic-speech evaluation.
  It finally builds a TTS corpus from dark data by utterance-wise filtering.%
}
\label{fig:3:dataflow}
\end{figure}

\section{Data Selection Based on Training Data Quality}
\label{sec:chapter3}
In this section, we describe a method for constructing a corpus from internet data using data selection based on training data quality. 
Our data selection is based on the \textit{training data quality}, which we define as the expected contribution of each utterance to the performance of a given TTS model. Unlike conventional data selection approaches that rely on predetermined, model-agnostic criteria (e.g., acoustic quality scores), our method evaluates how much each utterance improves the perceptual quality of the synthesized speech generated by the model.

An overview of the proposed method is shown in Fig.~\ref{fig:3:dataflow}. We perform data selection on speech data collected from YouTube using an evaluation-in-the-loop framework. In this loop, we train a TTS model with each candidate utterance and predict the perceptual naturalness of the resulting synthetic speech using a pseudo MOS model. Based on the pseudo MOS scores, we estimate the training data quality of each utterance. Finally, utterances with higher estimated quality are selected to construct the training corpus.

\subsection{Data Collection and Pre-Screening}
\label{sec:prescreening}
The first step is to collect text-audio pairs and pre-screen dark data to filter too low-quality data for TTS training.
We collect a dataset from YouTube and combine data screening methods for \donemodify{ASR}{automatic speech recognition (ASR)} and \donemodify{ASV}{automatic speaker verification (ASV)}.
We briefly describe each method below, but see the paper~\cite{takamichi2021jtubespeech} for more detail.

\subsubsection{Download Text-Audio Pairs from YouTube}
Following the methodology in the previous study\cite{takamichi2021jtubespeech}, we first compile a list of query terms, automatically extracted from two sources: (i) hyperlinked words appearing in Wikipedia articles, and (ii) trending keywords identified by Google Trends. Each term is then used to search YouTube, and only videos that include manually created subtitles are retained. For each selected video, we download both the audio track and its corresponding subtitle file, thereby obtaining time-aligned text-audio pairs.

\subsubsection{Pre-Screening via Text-Audio Alignment Accuracy}
Since TTS training requires that speech segments be reasonably aligned with their transcriptions, we use the connectionist temporal classification (CTC) score~\cite{kurzinger2020ctc} to assess the degree of alignment between audio and text. We apply CTC-based segmentation to split each audio file into utterances and compute alignment scores with respect to the corresponding YouTube subtitles.
Utterances with notably low scores are filtered out to avoid cases that could severely disrupt the following training process.

\subsubsection{Cleansing Based on Speaker Compactness}
For TTS training, it is preferable to have multiple consistent utterances per speaker. To assess speaker consistency, we compute the variance of $x$-vectors~\cite{snyder2018x} within each utterance group—defined as all utterances extracted from a single YouTube video. This variance serves as a proxy for the stability of speaker identity across the group. Groups exhibiting unusually high variance are filtered out, following the same principle as the alignment-based filtering.

After this step, each remaining group is assumed to contain utterances from a single, consistent speaker. It is worth noting that this method may inadvertently exclude some single-speaker groups with significant speaking-style variation, since $x$-vectors are sensitive to such variation even within the same speaker~\cite{williams2019disentangling}. Nevertheless, we adopt this approach because our goal is to capture speaker diversity rather than style diversity.

\subsection{Initial Training Using All the Candidate Data} 
We perform the initial training of the given TTS model using all the candidate data.  
To evaluate the quality of the training data for the desired TTS model, we consider it necessary to include a training step to capture the characteristics of the TTS model.  
In this step, we train a multi-speaker TTS model.  
Since the quality of synthesized speech is expected to vary across speakers in a multi-speaker TTS model, we infer the quality of the training data from the differences observed among speakers.

\begin{figure*}[t]
\highlight{
\centering
\includegraphics[width=0.98\linewidth]{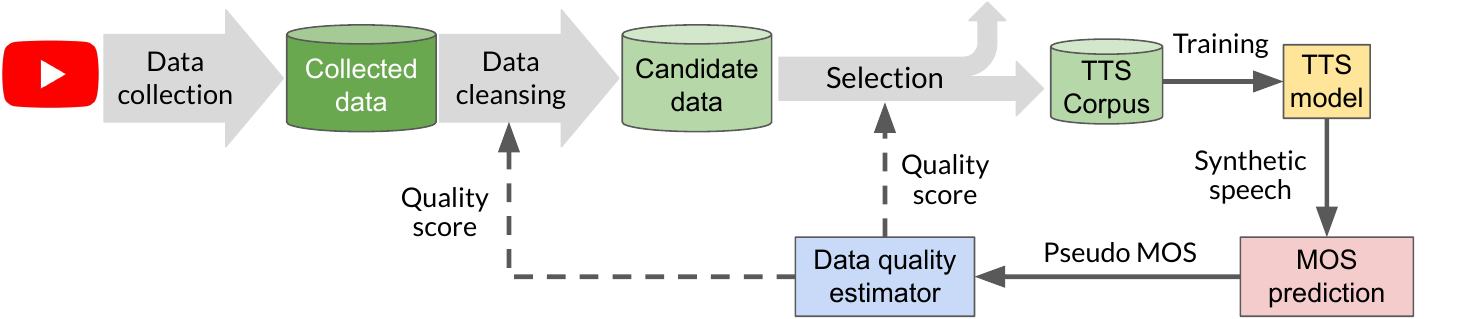}
\caption{
  Procedure of TTSOps.
  We obtain dark data from YouTube
    and evaluate each utterance through loop
      of TTS model training and synthetic-speech evaluation.
  It finally builds TTS corpus from dark data by utterance-wise filtering.%
}
\label{fig:4:overview}
}
\end{figure*}
\begin{figure}[t]
\centering
\includegraphics[width=1.0\linewidth]{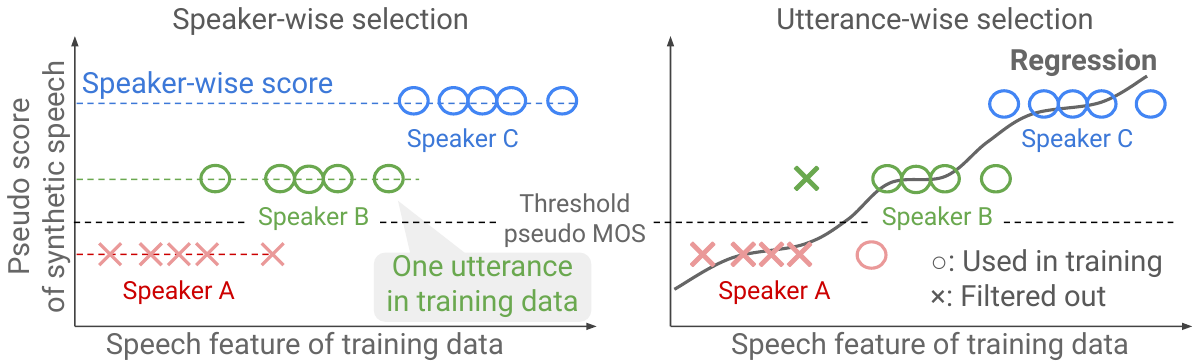}
 \caption{
   Comparison of speaker-wise and utterance-wise selection.
   With regression, we filter out low-score utterances even if speaker's pseudo MOS is high}
 \label{fig:reg}
\vspace{-10pt}
\end{figure}

\subsection{Evaluating Synthetic Speech Quality}
\label{section:3:regression}
We evaluate the quality of speech synthesized by the initially-trained TTS model.
A simple evaluation method is to calculate the value of the loss function for each utterance during training (e.g., the distance between the ground-truth and the predicted features).
However, the distance does not necessarily correspond to the perceptual quality of the synthesized speech~\cite{hayashi2021espnet2,weiss2021wave}.

Since MOS directly reflects perceptual quality, subjective evaluation is a natural choice.
However, it does not scale well when handling large volumes of data required for training.
To address this, we use pseudo MOS predicted by an automatic MOS prediction model.
These models have been reported to correlate well with human perceptual ratings~\cite{cooper2024review, huang2024voicemos}.
Therefore, we use the pseudo MOS score predicted by a pre-trained MOS prediction model as a proxy for naturalness in subjective evaluation.

This evaluation is used to determine the score of each training data, as described in the following Section~\ref{sec:3:utt_level}.
In other words, this evaluation aims to estimate the difference in the effect of each data on the quality of the synthesized speech.
The simplest method is to synthesize training data sentences and to filter out sentences with lower scores.
However, this method uses different sentences among speakers.
It is inappropriate because
   1) the pseudo MOS score changes depending on 
         the sentence to be synthesized~\cite{saeki2022utmos} and
   2) a sentence set greatly varies among speakers in dark data.
Therefore, we evaluate the quality of synthesized speech for each speaker, using common sentences.
In this way, it is possible to quantify the difference in synthetic speech quality of each speaker, without depending on the speakers' utterances.
Therefore, we evaluate the quality of synthesized speech for each speaker using common sentences not included in the training data, and averages of the values are used for each speaker.

\subsection{Quantifying Training Data Quality Score for Each Utterance}
\label{sec:3:utt_level}

We filter the training data on the basis of the obtained speaker-wise pseudo MOS.
The simplest method is to filter at the speaker level on the basis of the values, i.e., filtering out speakers with lower pseudo MOSs. 
However, since the data quality varies within the same speaker, filtering should be performed at the utterance level. 
In other words, we should filter out low-quality utterances even if speaker's pseudo MOS is high, and vice versa as shown in Fig.~\ref{fig:reg}.

For this purpose, we train a regression model that predicts the speaker-wise pseudo MOS from each utterance in the training data. 
We assume that acoustically similar training data will achieve similar naturalness in the synthesized speech. 
Furthermore, we assume that a regression model predicts close values for acoustically similar data.
These assumptions motivate us to use a regression model.

We train this regression model on all the pre-screened data.
We confirm the effect of using a regression model in the experiment described in Section~\ref{sec:PseudoMOSExperimentResult}.

\subsection{Training Data Selection and Re-Training} 
We evaluate the training data at the utterance level with the regression model.
Then, utterances with lower scores are filtered out.
Finally, we retrain the TTS model with the filtered data.

\begin{figure*}[t]
\highlight{
\centering
\begin{subfigure}[b]{0.4\linewidth} 
    \centering
    \includegraphics[width=1.0\linewidth]{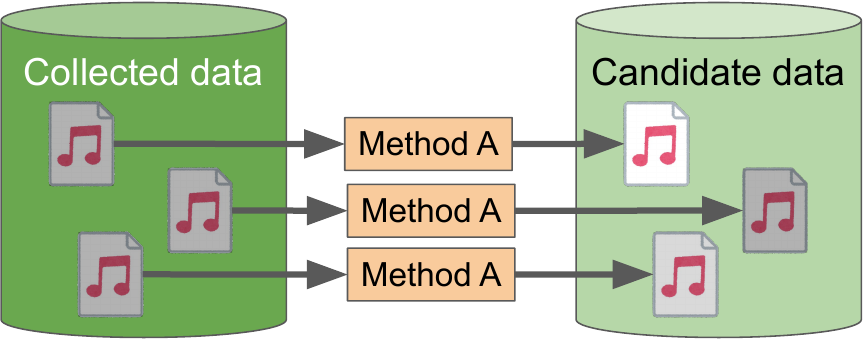}
    \caption{In conventional methods, the same data cleansing method is uniformly applied to all data.}
\end{subfigure}
\hspace{0.05\linewidth} 
\begin{subfigure}[b]{0.4\linewidth}
    \centering
    \includegraphics[width=1.0\linewidth]{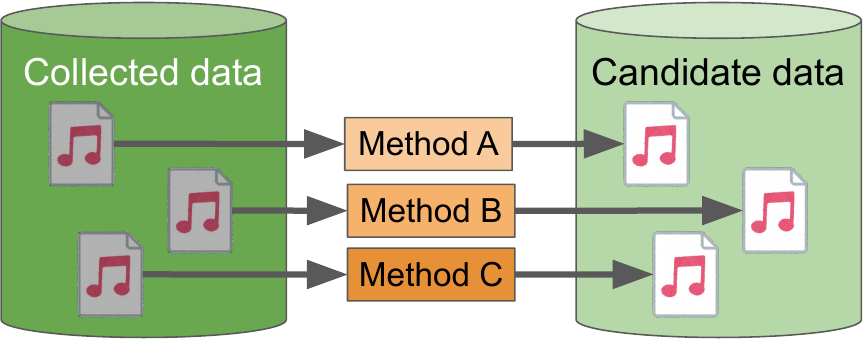}
    \caption{In the proposed method, different data cleansing methods are applied to each data point.}
\end{subfigure}
\caption{Comparison between conventional data cleansing procedures and the proposed data cleansing procedure.}
\label{fig:4:switch}
}
\end{figure*}

\section{TTSOps: Automatic Optimization Method of TTS Training Procedure}
\label{sec:chapter4}

In this section, we propose TTSOps, a unified training pipeline that extends the methodology presented in Section~\ref{sec:chapter3} by integrating data cleansing strategies into the evaluation-in-the-loop framework. As illustrated in Fig.~\ref{fig:4:overview}, TTSOps operates on speech data collected from the internet and estimates the training data quality—defined as each utterance's expected contribution to the performance of a TTS model—through a closed-loop process that combines data cleansing, data selection, model training, and pseudo MOS evaluation.

Unlike conventional pipelines that apply a fixed data cleansing method across the entire dataset, TTSOps performs utterance-wise adaptive switching of cleansing techniques, selecting the one that maximally improves the training data quality for each utterance as shown in Fig.~\ref{fig:4:switch}.
Based on this adaptive cleansing, the framework performs data selection guided by the estimated training data quality, ultimately constructing a high-quality corpus tailored to the characteristics of the target TTS model.

\subsection{Data Collection}
We collected the dataset following the method described in Section~\ref{sec:prescreening}.
Specifically, we extracted text-audio pairs from YouTube videos and applied a pre-screening process 
based on CTC alignment scores and speaker compactness using x-vectors.
This pre-screening was intended to prevent the initial training phase from failing by removing utterances 
with extremely low quality. For further details, refer to Section~\ref{sec:prescreening}.

\subsection{Training Data Quality Evaluation}
\begin{figure*}[t]
\highlight{
\centering
\includegraphics[width=0.85\linewidth]{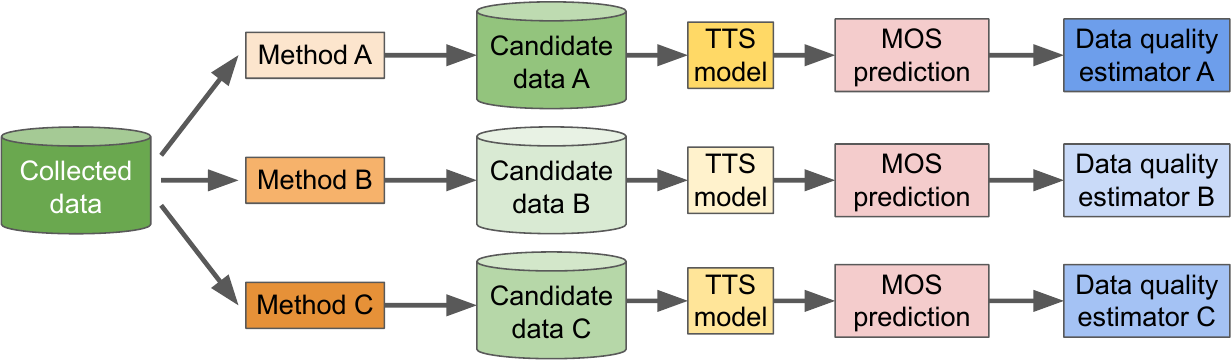}
\caption{
    A method for evaluating the quality of training data for each data cleansing method.  
    The training and evaluation loop is applied to datasets obtained by uniformly applying each data cleansing technique, and the training data quality is assessed.%
}
\label{fig:4:evaluate}
}
\end{figure*}

The method for evaluating the quality of training data for each data cleansing method is illustrated in Fig.~\ref{fig:4:evaluate}. 
In the proposed approach, each data cleansing method is uniformly applied to all data, and the quality of the training data is assessed through the training-evaluation loop described Section~\ref{sec:chapter3}.
This method involves training a TTS model using candidate data, evaluating the quality of the synthesized speech generated by the trained TTS model for each speaker, and learning a regression model to predict the quality of synthesized speech based on the training data. 
By predicting the synthesis quality of a TTS model trained with specific data, this framework evaluates the quality of the training data.

This method leverages the fact that, even with a single multi-speaker TTS model, the quality of synthesized speech varies among speakers. Therefore, it evaluates the quality of each dataset through a single instance of model training. However, when combining multiple data cleansing methods, as in this study, a single model training instance does not allow for separate evaluations of data cleansing methods. 
To address this limitation, this study evaluates the quality of training data separately for each data cleansing method.

\subsection{Corpus Determination}

\begin{figure}[t]
\centering
\highlight{
\includegraphics[width=0.95\linewidth]{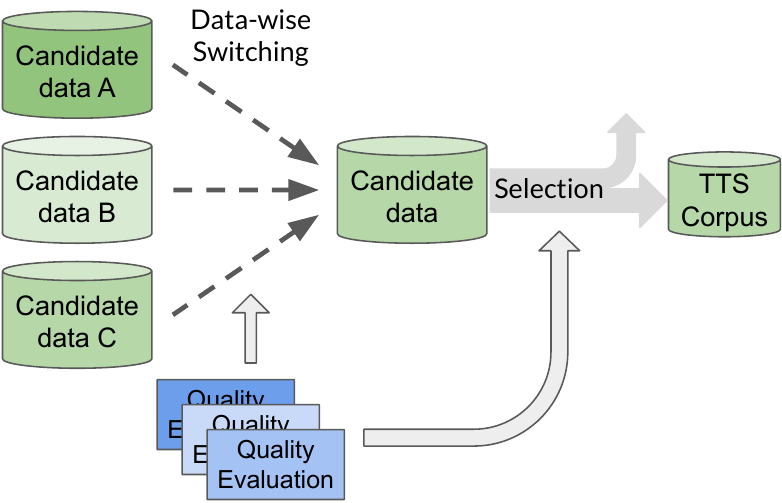}
\caption{
    Corpus determination procedure.  
    Preprocessing methods are switched data-wise based on training data quality, and the final corpus is determined through data selection based on the training data quality of each dataset.%
}
\label{fig:4:cleans_and_select}
}
\end{figure}

To determine final corpus, data cleansing and selection are performed using the method shown in Fig.~\ref{fig:4:cleans_and_select}, based on training data quality evaluation.  
In the data cleansing process, the data cleansing method that results in the highest training data quality is selected and applied for each dataset.  
Furthermore, since the training data quality has already been evaluated using the method described in Section~\ref{sec:chapter3}, data selection is conducted accordingly.  

Here, by saving the preprocessed data, the need to repeat data cleansing is eliminated, enabling data cleansing selection simply by gathering the relevant files.  
Specifically, the training data quality obtained from the adopted data cleansing method is directly regarded as the quality of each dataset. Data with higher quality is selected in descending order to construct the TTS corpus.  
The constructed corpus is then used to train the TTS model.

\section{Experimental Evaluation of Data Selection}
\label{sec:experiment_data_selection}
\subsection{Experimental Setting}
\subsubsection{Dataset}
\label{sec:3:dataset}
We followed JTubeSpeech~\cite{takamichi2021jtubespeech} scripts\footnote{\url{https://github.com/sarulab-speech/jtubespeech}} to obtain dark data from YouTube; the amount was approximately $3{,}500$ hours.
Pre-screening with a CTC threshold of $-0.3$ and speaker compactness threshold of $[1, 7]$\footnote{These values are the same as the experiments in the JTubeSpeech paper~\cite{takamichi2021jtubespeech}.}, we obtained approximately $66$~hours ($60,000$~Japanese utterances) of $2,719$ speakers as the pre-screened data.
The sentences used for calculating the pseudo MOS were $100$ phoneme-balanced sentences from the JVS corpus~\cite{takamichi2019jvs}. 
The test data used to evaluate the finally trained TTS models
   was $324$ sentences from the ITA corpus\footnote{\url{https://github.com/mmorise/ita-corpus}}. 
There was no overlap in text among the pre-screened data,
  sentences for the pseudo MOS, and test data.

\subsubsection{Model and Training}
\label{sec:ModelAndTraining}
We trained an acoustic model and combined it with the pre-trained HiFi-GAN vocoder~\cite{kong2020hifi} UNIVERSAL\_V1\footnote{\url{https://github.com/jik876/hifi-gan}} as a multi-speaker TTS system.
We used two acoustic models, FastSpeech~2~\cite{ren2020fastspeech} and \donemodify{GlowTTS~\cite{kim2020glow}}{Matcha-TTS~\cite{mehta2024matcha}}.
We followed the model size and hyperparameters of the open-sourced implementation\footnote{FastSpeech~2: \url{https://github.com/Wataru-Nakata/FastSpeech2-JSUT}}\donemodify{\footnote{GlowTTS: \url{https://github.com/trgkpc/glow-tts_JSUT}}}{\footnote{Matcha-TTS:\url{https://github.com/shivammehta25/Matcha-TTS}}} except for the speaker representation.
Instead of the one-hot speaker representation implemented in the repository, we used an open-sourced $x$-vector extractor\footnote{\url{https://github.com/sarulab-speech/xvector_jtubespeech}}, and a $512$-dimensional $x$-vector was used to condition the TTS model. 
The $x$-vector was added to the output of the FastSpeech~2 encoder via a $512$-by-$256$ linear layer. 
The $x$-vector was averaged for each speaker; one $x$-vector corresponded to one unique speaker.
The TTS model was pre-trained using $10,000$ utterances from the JVS corpus~\cite{takamichi2019jvs}, the $100$-speaker Japanese TTS corpus. 
We performed 300k steps with a batch size of $16$ in this pre-training.
TTS training in this paper started from this pre-trained model with 100k steps with a batch size of $16$.

We used a pre-trained UTMOS~\cite{saeki2022utmos} strong learner\footnote{\scriptsize\url{https://github.com/sarulab-speech/UTMOS22}} to obtain a five-scale pseudo MOS on naturalness from synthetic speech. 
The regression model for predicting the pseudo MOS from the training data was $1$-layer $256$-unit bi-directional long short-term memory~\cite{hochreiter1997long}, followed by a linear layer, ReLU activation, and another linear layer. 
We used frame-level self-supervised learning~(SSL) features\footnote{
   We compared the SSL features and the NISQA~\cite{mittag2021nisqa} features (used in the baseline) in the preliminary evaluation. The result demonstrated that the SSL features performed better.
  } obtained with a wav2vec~2.0 model~\cite{baevski2020wav2vec}\footnote{\scriptsize\url{https://github.com/facebookresearch/fairseq/tree/main/examples/wav2vec}}, as the input.
The frame-level outputs were aggregated to predict the pseudo MOS.
The number of training steps, minibatch size, optimizer, and training objective were 10k, $12$, Adam~\cite{kingma2014adam} with a learning rate of $0.0001$, and mean squared error, respectively. 

\subsubsection{Compared Methods}
\label{sec:3:compared}
We compared the following data selection methods.
\begin{itemize}\leftskip -3mm \itemsep 0mm
    \item \textbf{Unselected}: 
    All the pre-screened data was used for the TTS training; 
      the training data size was approximately $60,000$ utterances. 
    \item \textbf{Acoustic-quality~(utterance-wise)}: 
    The training data was selected in terms of the acoustic quality of the data. 
    We used NISQA~\cite{mittag2021nisqa}, a recent deep learning-based model to predict scores on 
      naturalness, noisiness, coloration, discontinuity, and loudness of the speech data. 
    Each score takes $[1,5]$.
    We chose multiple threshold $\theta=4.0,3.5,3.0,2.5$.
    That is data for which all the scores were higher than $\theta$ were selected. 
    For $\theta=4.0,3.5,3.0,2.5$, the TTS training data size $n$ was $5128, 12172, 20390$ and $31038$, respectively.
   \item \textbf{Ours-Utt~(evaluation-in-the-loop utterance-wise selection)}: 
   Our evaluation-in-the-loop data selection. 
   For each data, 
     we estimate the speech quality synthesized 
     by the TTS model from an initial training 
       with pre-screened data.
   The threshold for selecting training data was set to 
     have the resulting training data be the same in size as ``Acoustic-quality''.
   \item \textbf{Ours-Spk~(evaluation-in-the-loop speaker-wise selection)}:
   Our data selection, but the data selection was performed per speaker as described in Section~\ref{section:3:regression}. 
   The threshold for selecting training data was set to 
     have the resulting in the size of the training data be almost the same as ``Acoustic-quality''.
\end{itemize}
\subsubsection{Evaluation}
\label{sec:3:eval}
We evaluated the selection methods to clarify the following:
\begin{itemize}\leftskip -3mm \itemsep 0mm
    \item \textbf{Does our method obtain more ``high-quality speakers?'': pseudo MOS comparison.} Our TTS model is expected to reproduce voices for a higher number of speakers. We define a ``high-quality speaker'' as a speaker with a higher pseudo MOS than the threshold. 
    We in advance trained an high-quality multi-speaker TTS model
      using the JVS corpus~\cite{takamichi2019jvs}
      and calculated the speaker-wise pseudo MOS scores.
    We set the lowest score among the JVS speakers as the threshold\footnote{
        The JVS corpus was constructed in a well-designed environment, 
          and we confirmed that it was not an outlier.
    }.
    Speakers with a higher score than the threshold were considered to be high-quality speakers. 
    For each data selection method, we calculated
      1) the distribution of the pseudo MOSs for synthetic speech by the trained TTS model
        and 
      2) the number of high-quality speakers. 
    \item \textbf{Does our method increase speaker variation?} 
    We evaluated whether our method obtains diverse (i.e., sounding different) high-quality speakers. 
    To quantify the speaker variation, we calculated the cost of a Euclidean minimum spanning tree~\cite{march2010fast} of $x$-vectors of the high-quality speakers. 
    The calculation is similar to the g2g (median of the distances to the nearest $x$-vector)~\cite{stanton2022speaker},
     but we used summation instead of a representative value~(i.e., median) 
     because our purpose is to evaluate how widely speakers spread.
     \item \textbf{Does synthetic speech of so-called ``high-quality speakers'' truly sound natural?: actual MOS evaluation.} 
    We evaluated whether our data selection based on pseudo MOS is truly effective in synthesizing perceptually natural speech. 
    To this end, we conducted a subjective evaluation of synthetic speech quality using different data selection methods, and investigated their performance in relation to pseudo MOS.    
\end{itemize}

Note that seen and unseen speakers are different among the data selection methods. 
All the speakers were seen ones for ``Unselected,'' but only parts of them were seen in the other methods. Also, seen speakers were different among ``Acoustic-quality,'' ``Ours-Utt,'' and ``Ours-Spk.'' 
Speakers not used for training 
  were considered to be unseen speakers for each method.
Unless otherwise noted, we describe results aggregating those of both seen and unseen speakers.

\subsection{Results}
\newcommand{\normalsinglepseudomos}[4]{
    \begin{subfigure}{1.0\linewidth}
    \includegraphics[width=1.0\linewidth]{chapter3/img/#1/normal_pseudo_mos_Uncleansed.pdf}
    \vspace{#3}
    \caption{#2}
    \vspace{#4}
    \end{subfigure}
}

\begin{figure}[t]
  \centering
  
  \normalsinglepseudomos{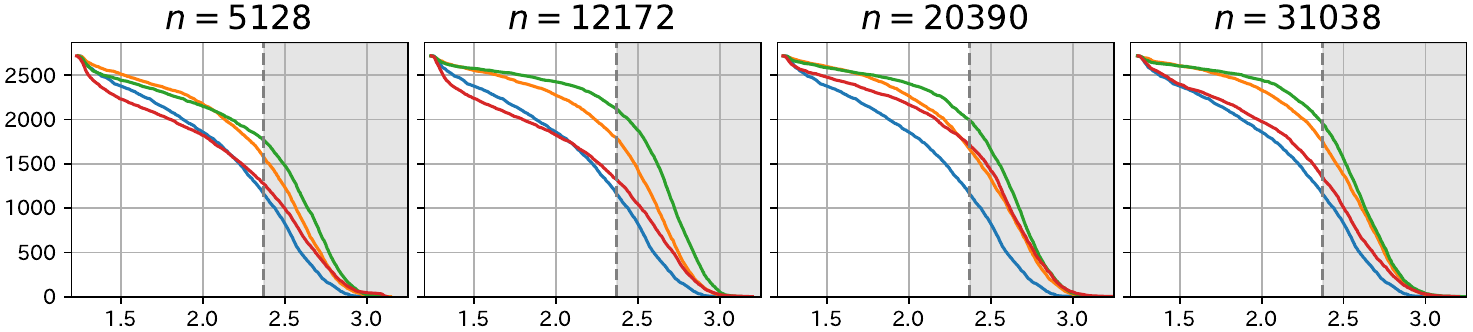}{FastSpeech~2}{-13pt}{5pt}
  \donehighlight{
  \normalsinglepseudomos{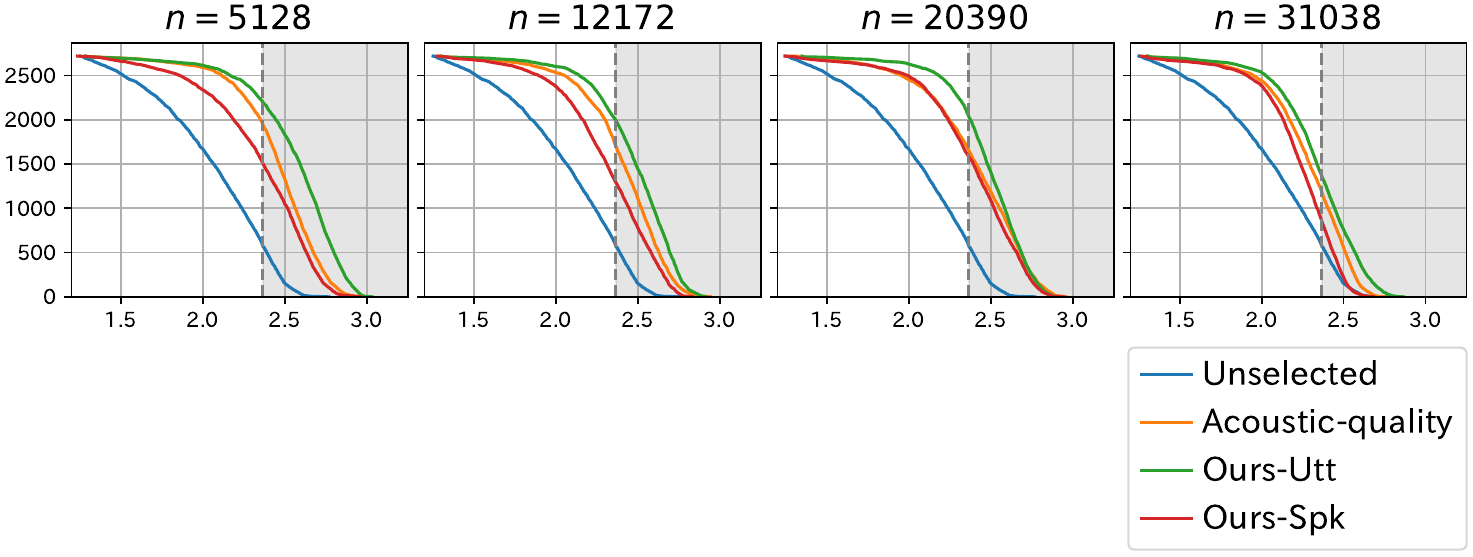}{\donemodify{GlowTTS}{Matcha-TTS}}{-45pt}{10pt}
  }

  \caption{
    Cumulative histograms of pseudo MOS. 
    Y-axis value indicates number of speakers with higher score than x-axis value.
    The shaded area corresponds to high-quality speakers.
  }
  \label{fig:3:pseudoMOS}
\end{figure}

\begin{table}[t]
  \caption{
    Number of overall, seen and unseen speakers. 
    Values of each cell are number of high-quality speakers, all speakers, and ratio of two values, respectively. 
  }
  \centering

\footnotesize
\begin{tabular}{c|c||c|c}
$n$ & Method & FastSpeech~2 & \donemodify{GlowTTS}{Matcha-TTS} \\ \midrule \midrule    
\multirow{1}{*}{$58500$}
  &Unselected & $1157 ~(42.6\%)$ & \donemodify{}{$578 ~ (21.3\%)$} \\ \midrule
\multirow{3}{*}{$5128$}
  &Acoustic-quality & $1572 ~(57.8\%)$ & \donemodify{}{$1949 ~ (71.7\%)$} \\
  &Ours-Utt & $\mathbf{1764 ~(64.9\%)}$ & \donemodify{}{$\mathbf{2198 ~ (80.8\%)}$} \\
  &Ours-Spk & $1274 ~(46.9\%)$ & \donemodify{}{$1505 ~ (55.4\%)$} \\ \midrule
\multirow{3}{*}{$12172$}
  &Acoustic-quality & $1786 ~(65.7\%)$ & \donemodify{}{$1698 ~ (62.4\%)$} \\
  &Ours-Utt & $\mathbf{2114 ~(77.7\%)}$ & \donemodify{}{$\mathbf{1993 ~ (73.3\%)}$} \\
  &Ours-Spk & $1321 ~(48.6\%)$ & \donemodify{}{$1299 ~ (47.8\%)$} \\ \midrule
\multirow{3}{*}{$20390$}
  &Acoustic-quality & $1661 ~(61.1\%)$ & \donemodify{}{$1654 ~ (60.8\%)$} \\
  &Ours-Utt & $\mathbf{1989 ~(73.2\%)}$ & \donemodify{}{$\mathbf{2035 ~ (74.8\%)}$} \\
  &Ours-Spk & $1712 ~(63.0\%)$ & \donemodify{}{$1583 ~ (58.2\%)$} \\ \midrule
\multirow{3}{*}{$31038$}
  &Acoustic-quality & $1748 ~(64.3\%)$ & \donemodify{}{$1188 ~ (43.7\%)$} \\
  &Ours-Utt & $\mathbf{1955 ~(71.9\%)}$ & \donemodify{}{$\mathbf{1361 ~ (50.1\%)}$} \\
  &Ours-Spk & $1347 ~(49.5\%)$  & \donemodify{}{$853 ~ (31.4\%)$}
\end{tabular}
\end{table}

\newcommand{\hq}[4]{
    \begin{subfigure}{1.0\linewidth}
    \includegraphics[width=1.0\linewidth]{chapter3/img/#1/hq_350.png}
    \vspace{#3}
    \caption{#2}
    \vspace{#4}
    \end{subfigure}
}

\begin{figure}[t]
  \centering
  
  \hq{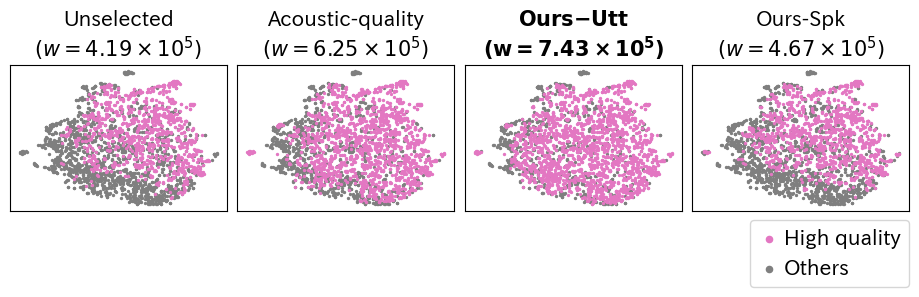}{FastSpeech~2}{-35pt}{10pt}
  \donehighlight{
  \hq{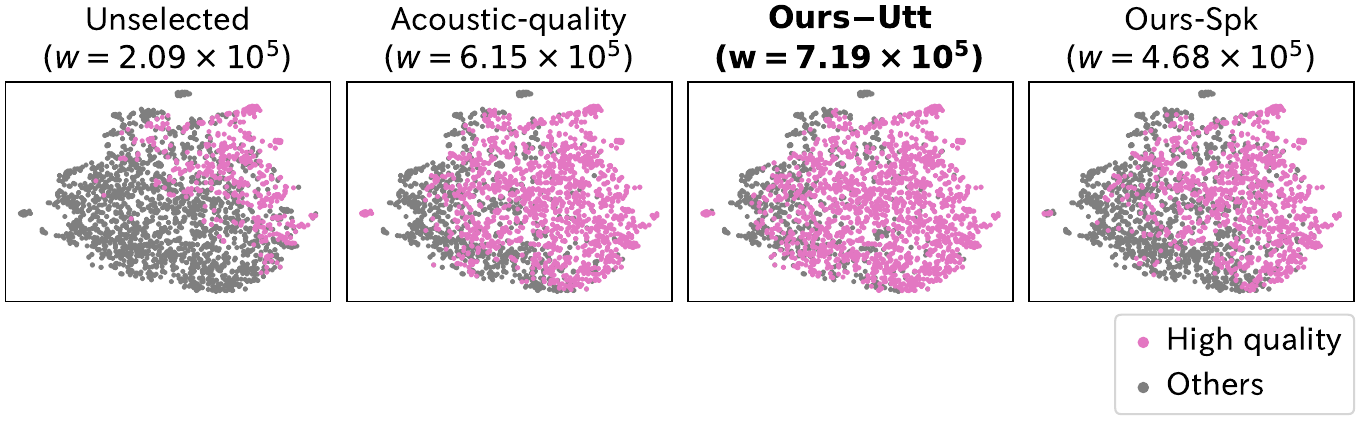}{\donemodify{GlowTTS}{Matcha-TTS}}{-32pt}{-10pt}
  }
  
  \caption{
    Distributions of high-quality speakers by each data selection method~($n{=}12172)$.
    Higher $w$ indicates more diversity.
  }
  \label{fig:3:hq}
\end{figure}

\subsubsection{Number of High-Quality Speakers}
\label{sec:PseudoMOSExperimentResult}

Fig.~\ref{fig:3:pseudoMOS} is a cumulative histogram of the pseudo MOSs. 
Our method had the highest values among the methods, demonstrating that a TTS model trained on our data selection can synthesize multi-speaker voices with higher quality than the other methods. 
The numbers of high-quality speakers for ``Unselected,'' ``Acoustic-quality,'' ``Ours-Utt,'' and ``Ours-Spk'' were $924, 1737, 1942$, and $1367$, respectively.
We see that the proposed method increased the number of high-quality speakers, compared with the other methods. 
Specifically, the increment was approximately $1.2$ times from that of ``Acoustic-quality.'' 
From these results, we can say that the proposed method work better than the conventional methods for the purpose of increasing the speaker variety of the multi-speaker TTS model.
Note that this TTS corpus is much larger than the previous Japanese multi-TTS corpus~(JVS) composed of 100 speakers.

In addition, comparing ``Ours-Utt'' and ``Ours-Spk'', the proposed method was significantly better in terms of both the pseudo MOS distribution and the number of the high-quality speaker. 
This indicates that utterance-wise selection significantly contributes to enhancing the performance, rather than speaker-wise selection.

\subsubsection{Evaluation of Speaker Variation}
To validate the effectiveness of dark data, Fig.~\ref{fig:3:hq} shows the distributions of the high-quality speakers and speaker variation scores $w$. 
Both qualitatively and quantitatively, ``Ours-Utt'' increased the speaker variation compared with the other methods, indicating that our method contributes to speaker variation.

\begin{figure}[t]
\centering
\donehighlight{
\includegraphics[width=1.0\linewidth]{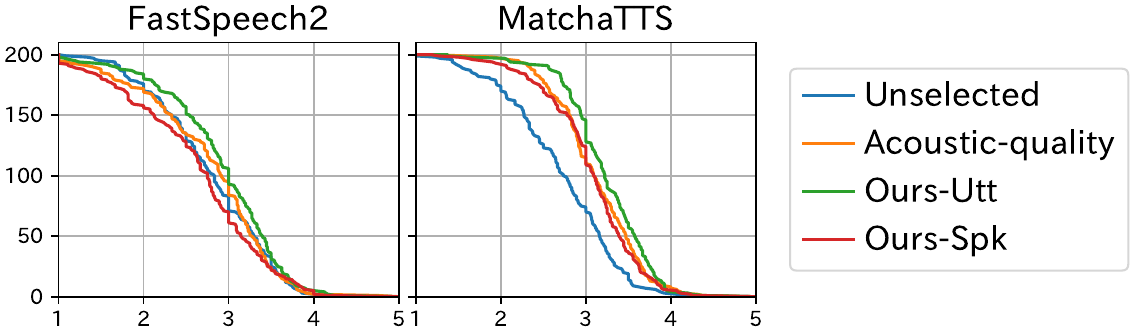}
  \caption{
    Cumulative histograms of actual MOS. 
    Y-axis value indicates number of speakers with higher score than x-axis value.
  }
\label{fig:3:subjective}
}
\end{figure}
\newcommand{\compareactualvspseudoeach}[3]{
    \begin{subfigure}{1.0\linewidth}
    \includegraphics[width=1.0\linewidth]{chapter#1/img/compare/compare_#2.pdf}
    \vspace{-15pt}
    \caption{#3}
    \vspace{5pt}
    \end{subfigure}
}

\newcommand{\compareactualvspseudo}[1]{
\begin{figure}[t]
  \centering
  
  \compareactualvspseudoeach{#1}{fs2}{FastSpeech~2}
  \compareactualvspseudoeach{#1}{glow}{GlowTTS}
  \vspace{-18pt}
  
  \caption{
    Comparison between pseudo MOS and actual MOS. Each point corresponds to an individual speaker.
  }
  \label{fig:#1:mos_vs_pseudo_mos}
  \vspace{-5pt}
\end{figure}    
}
\begin{figure}[t]
\centering
\donehighlight{
\includegraphics[width=0.9\linewidth]{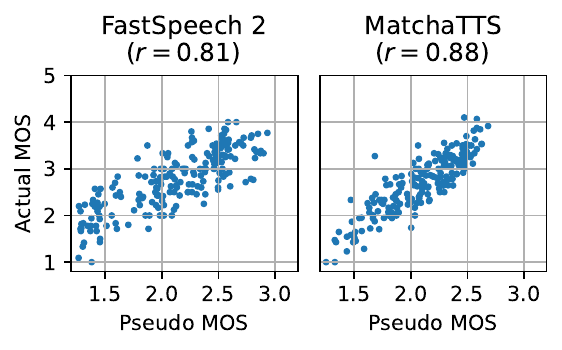}
\vspace{-5pt}
  \caption{
    Scatter plots showing the relationship between pseudo MOS and actual MOS for each speaker.     
    \donemodify{Although a consistent correlation is observed in both models, the correlation is weaker for GlowTTS, leading to reduced effectiveness of the proposed method.}{
    A strong correlation was observed between pseudo MOS and actual MOS across speakers ($r > 0.8$).
    }
  }
\label{fig:3:uncleansed}
}
\end{figure}
\begin{figure}[t]
\donehighlight{
\centering
   \includegraphics[width=1.0\linewidth]{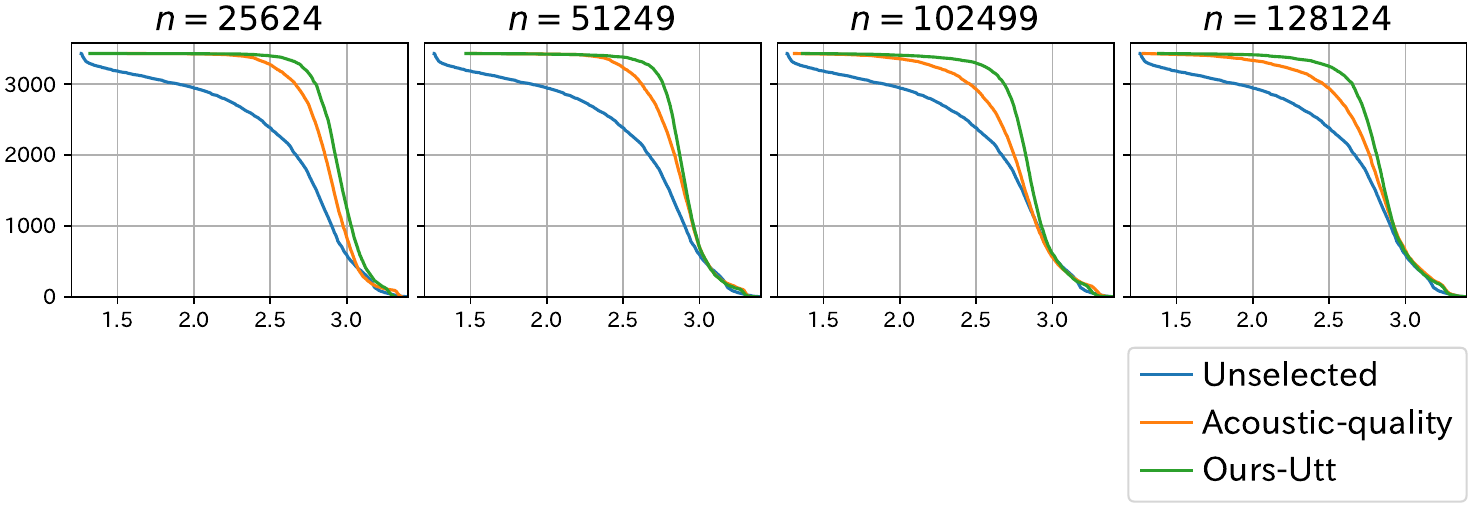}  
  \caption{
  \donemodify{}{
    Evaluation results of the additional experiment on English data. Cumulative histograms of pseudo MOS.
  }}
\label{fig:3:giga}
}
\end{figure}

\subsubsection{Validation on Actual MOS}
\label{sec:mos_vs_pseudomos}
We conducted a five-point MOS test to assess the naturalness of synthetic speech. 
A total of $500$ listeners participated, each listening to $20$ samples.
For both FastSpeech~2 and \donemodify{GlowTTS}{Matcha-TTS}, we evaluated the synthetic speech generated by TTS models trained on data selected with each data selection method.
For ``Acoustic-quality,'' ``Ours-Utt,'' and ``Ours-Spk,'' the data set size $n$ was set to $12,172$.
To reduce the evaluation cost, we randomly sampled $200$ speakers from the total of $2,719$ for evaluation. 
The actual MOS scores were then aggregated for each speaker.

Fig.~\ref{fig:3:subjective} shows the evaluation results. 
In both models, ``Ours-Utt'' achieved the highest scores among all methods, indicating that the proposed method is more effective than conventional approaches even in terms of perceptual speech quality.

\donemodify{
However, the improvement observed in GlowTTS was less pronounced than in FastSpeech~2. 
To investigate the cause, we examined the relationship between pseudo MOS and actual MOS on ``Unselected.''
Fig.~\ref{fig:3:uncleansed} presents the scatter plots and the correlation coefficients $r$. 
While FastSpeech~2 showed a high correlation, GlowTTS exhibited lower correlation, possibly because the narrower distribution range of pseudo MOS in GlowTTS amplified the impact of prediction errors.
Since our proposed method estimates training data quality using the pseudo MOS scores from ``Unselected,'' its effectiveness in actual MOS is likely reduced in the case of GlowTTS, where pseudo MOS shows a lower correlation with actual MOS.
The limitations of the proposed method arising from the accuracy of pseudo MOS are discussed in Section~\ref{sec:mos_lim}.
}{
To investigate the reliability of pseudo MOS as an indicator of perceptual quality,
we compared the results of pseudo MOS and actual MOS on ``Unselected.''
Fig.~\ref{fig:3:uncleansed} presents the scatter plots and the correlation coefficients $r$.
\modify{
Both FastSpeech~2 ($r{=}0.81$) and Matcha-TTS ($r{=}0.88$) exhibited strong correlations,
indicating that pseudo MOS is a trustworthy predictor of actual MOS across different architectures.
}{
The correlation coefficient $r$ was $0.81$ for FastSpeech~2 and $0.88$ for Matcha-TTS, 
with the 95\% confidence intervals calculated using Fisher transformation~\cite{fisher1915frequency} being $[0.76, 0.85]$ and $[0.84, 0.91]$, respectively.
These results indicate that pseudo MOS is a trustworthy predictor of actual MOS across different architectures.
}
Since our proposed method estimates training data quality based on pseudo MOS,
this high correlation explains why the method’s effectiveness is also reflected in improvements of actual MOS.
The limitations of relying on pseudo MOS prediction accuracy are discussed in Section~\ref{sec:mos_lim}.
}

\donemodify{}{
\subsection{Evaluation of Cross-Lingual Generalization}
\label{sec:3:language}
To investigate the language-independence of the proposed framework, 
we further conducted experiments on English data. 
We used the \textit{L} subset of the GigaSpeech corpus~\cite{chen2021gigaspeech}, which contains 
diverse speech including YouTube data. 
After pre-screening based on speaker compactness, 
we obtained $3434$ speakers, $256{,}249$ utterances, and 
approximately $300$ hours of data.

We compared three data selection methods: 
``Unselected'', ``Acoustic-quality'', and ``Ours''. 
For a fair comparison, we constructed training sets of four different sizes, 
$n=25,624$, $51,249$, $102,499$, and $128,124$, 
corresponding to $10\%$, $20\%$, $40\%$, and $50\%$ of the entire pre-screened corpus. 
We used FastSpeech~2 as the TTS model for these experiments.

Figure~\ref{fig:3:giga} shows the cumulative distributions of pseudo MOS 
across different corpus sizes. 
In all conditions, the proposed method outperformed 
the acoustic-quality-based baseline, indicating that 
evaluation-in-the-loop selection is effective not only for Japanese 
but also for English. 
These results support the language-agnostic nature of the proposed framework.
}

\donemodify{}{
\begin{table}[t]
\donehighlight{
\centering
\caption{Measured computation time on NVIDIA H200.}
\label{table:time}

\begin{tabular}{l|c|c}
\hline
 & FastSpeech~2 & Matcha-TTS \\
\hline
Training    & 1h 26m 27s & 5h 9m 33s \\
Evaluation  & 1h 11m 36s & 2h 1m 14s \\
Regression  & 0h 9m 7s  & 0h 9m 7s \\
\hline
\textbf{Total} & \modify{\textbf{4h 13m 36s}}{\textbf{4h 13m 37s} } & \modify{\textbf{12h 29m 29s}}{\textbf{12h 29m 27s}} \\
\hline
\end{tabular}

}
\end{table}

\subsection{Computation Time Analysis}
\label{sec:3:cost}
We measured the actual computation time of our framework 
on an NVIDIA H200 GPU environment. 
The results are shown in Table~\ref{table:time}.
``Training'' denotes the training phase, ``Evaluation'' refers to the synthesis and evaluation phase, 
and ``Regression'' indicates the training and inference of the data quality estimator.
The total time was calculated as $2\times$\textit{Training} + \textit{Evaluation} + \textit{Regression}, 
since our method involves two training phases (initial training and retraining after data selection).

As a result, FastSpeech~2 finished in about 4 hours and 16 minutes in total. 
More than half of the total time was occupied by the ``Training'' phase, suggesting that accelerating training is particularly important. 
In addition, although less than training, the ``Evaluation'' phase still required a substantial amount of time, 
as it involves both TTS inference and MOS prediction. 
Matcha-TTS required more time for both training and inference, resulting in a total of about 12 and a half hours. 
In both cases, these execution times can be regarded as falling within a practically feasible range for real-world TTS development.

}

\newcommand{\samplingpseudomos}[4]{
    \begin{subfigure}{1.0\linewidth}
    \includegraphics[width=1.0\linewidth]{chapter3/img/#1/sampling.pdf}
    \vspace{#3}
    \caption{#2}
    \vspace{#4}
    \end{subfigure}
}

\begin{figure}[t]
  \centering
  
  \donehighlight{
  \samplingpseudomos{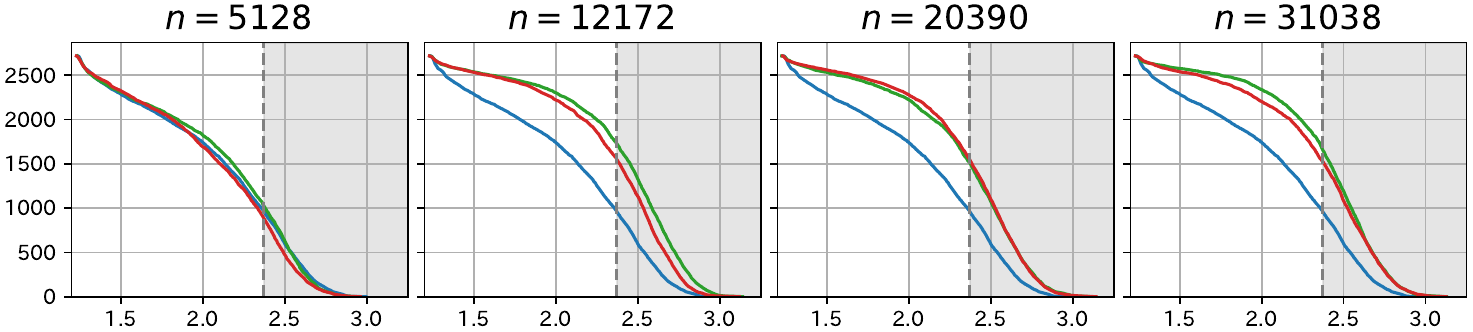}{\donemodify{}{FastSpeech~2}}{-13pt}{5pt}
  \samplingpseudomos{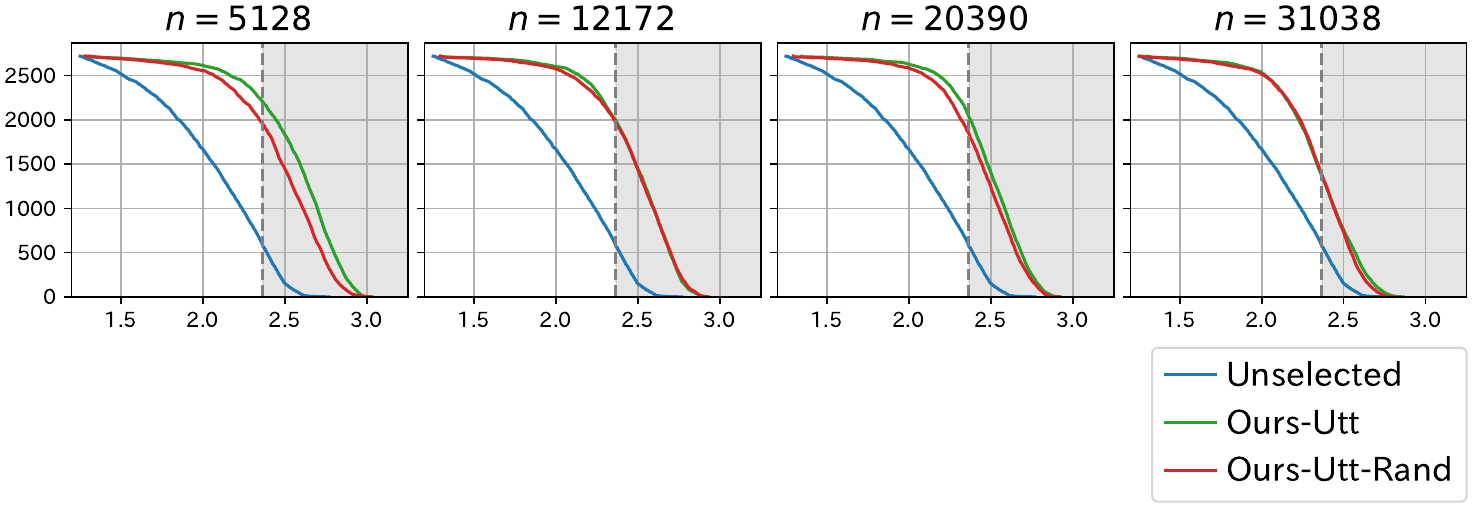}{\donemodify{}{Matcha-TTS}}{-45pt}{8pt}

  \caption{\donemodify{}{
    Cumulative histograms of pseudo MOS. 
  }}
  \label{fig:3:sampling}
  }
  \vspace{-5pt}
\end{figure}

\donemodify{}{
\subsection{Sampling Strategy}
\label{sec:3:sampling}
To mitigate the computational cost, which is an inherent limitation of our framework, 
we further investigated whether the amount of data used 
for the initial training step can be reduced by sampling. 
Although our method requires training on the entire candidate set 
in the first iteration, the training data quality estimator 
does not necessarily need to see all utterances, 
as long as the sampled subset adequately represents the distribution. 
If this assumption holds, the regression model can still be trained effectively 
while significantly reducing the computational load.

To evaluate this idea, we introduced a variant named ``Ours-Utt-Rand'', 
in which only 10\% of the candidate data were used for the initial training 
to train the data quality estimator. 
The estimator was then applied to all candidate data, 
meaning that 90\% of the utterances were evaluated in an unseen manner. 
We compared this approach with the standard ``Ours-Utt''.

The results are shown in Fig.~\ref{fig:3:sampling}.
``Ours-Utt-Rand'' exhibited slightly lower performance than ``Ours-Utt'' 
under some conditions, 
but overall achieved comparable results. 
This indicates that, when scaling up to larger corpora, 
sampling strategies can be employed to reduce computational cost 
while retaining most of the benefits of the proposed method.
}

\begin{figure}[t]
\centering
\donehighlight{
\includegraphics[width=0.9\linewidth]{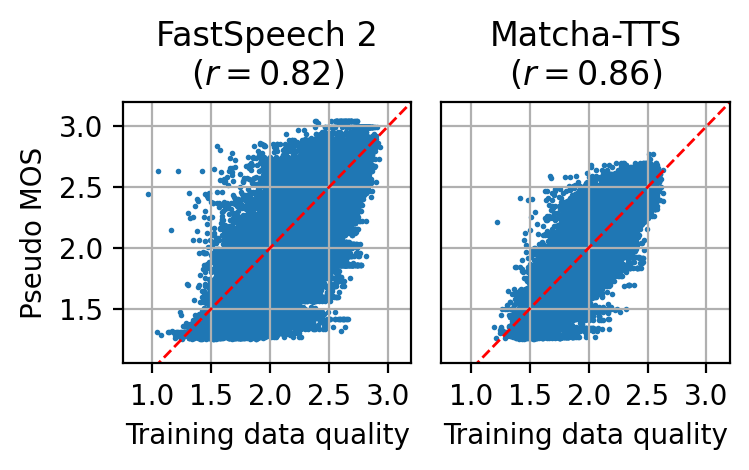}
  \caption{
    \donemodify{}{Scatter plots showing the relationship between pseudo MOS and training data quality.}
    }
\label{fig:3:tq_analysis}
}
\end{figure}

\newcommand{\aqanalysisfigure}[2]{
    \begin{subfigure}{1.0\linewidth}
    \includegraphics[width=0.9\linewidth]{chapter3/img/regression/data/#1.png}
    \vspace{-2pt}
    \caption{\donemodify{}{#2}}
    \vspace{5pt}
    \end{subfigure}
}

\begin{figure}[t]  
\donehighlight{
  \centering
  \aqanalysisfigure{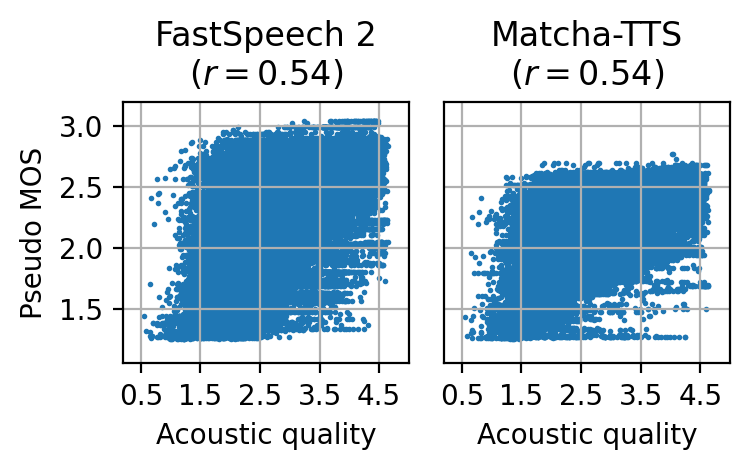}{Comparison with pseudo MOS.}
  \aqanalysisfigure{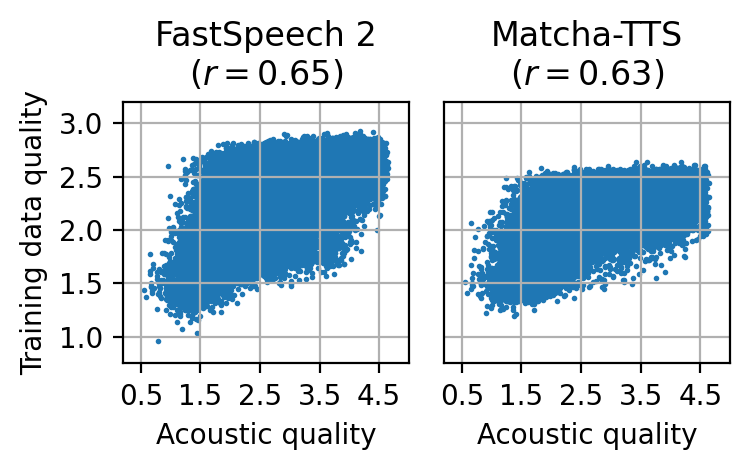}{Comparison with training data quality.}
  
  \vspace{-15pt}
  \caption{\donemodify{}{Correlation analysis of acoustic quality with other metrics.
  }}
  \label{fig:3:aq_analysis}
}
\end{figure}

\donemodify{}{
\subsection{Analysis of Data Quality Estimator}
To investigate the behavior of the data quality estimator, 
we analyzed the estimated training data quality. 
As shown in Fig.~\ref{fig:3:tq_analysis}, training data quality exhibited a strong correlation with pseudo MOS. 
Nevertheless, many utterances were scattered away from the diagonal. 
Horizontally, each spread indicates how utterances with the same pseudo MOS 
were distributed in terms of training data quality, and in some regions this spread exceeded $1.0$,
confirming that training data quality does not collapse to a single value even within the same pseudo MOS group.

To further examine the difference from acoustic quality, 
we analyzed acoustic quality in relation to other metrics. 
Fig.~\ref{fig:3:aq_analysis} presents the correlations of acoustic quality 
with pseudo MOS and with training data quality. 
The results show that acoustic quality exhibits only weak correlation with pseudo MOS, 
whereas its correlation with training data quality is relatively higher. 
This may be because acoustic quality constitutes a partial component 
of the comprehensive measure represented by training data quality, 
and therefore shows correlation with it.

Taken together, these findings suggest that training data quality 
possesses intermediate characteristics between pseudo MOS and acoustic quality. 
It retains correlation with the perceptual quality of synthetic speech, 
while simultaneously reflecting utterance-level characteristics of the training data.

}
\section{Experimental Evaluation of TTSOps}
\label{sec:experiment_ttsops}
\subsection{Experimental Setting}
\subsubsection{Dataset, Model and Training}
We utilized the same data collected from YouTube as described in Section~\ref{sec:3:dataset}.
Also, we utilized the same model and the same training method as described in Section~\ref{sec:ModelAndTraining}.

\subsubsection{Compared Methods}
\label{sec:4:exp:cleansing}
As data cleansing methods, the following were considered:  
\begin{itemize} \leftskip -3mm \itemsep -0mm
    \item \textbf{Uncleansed}:
        The data was used as is. This method is expected to be suitable for high-quality audio, such as studio-recorded speech.
    \item \textbf{Demucs}:
        As a preprocessing method, we applied a pretrained speech enhancement model, Demucs~\cite{rouard2022hybrid}, to all data\footnote{\url{https://huggingface.co/spaces/Wataru/Miipher/tree/main}}.
        Demucs is a type of source separation model trained to remove background music. 
        Therefore, it is expected to reduce additive noise and improve data quality.
    \item \textbf{Miipher}:
        As a preprocessing method, we applied a pretrained audio restoration model, Miipher~\cite{koizumi2023miipher}, to all data\footnote{\url{https://github.com/facebookresearch/demucs}}.
        This model is trained to restore degraded audio. 
        Consequently, it is expected to reduce environmental distortions caused by recording devices or reverberation, thereby enhancing data quality.
    \item \textbf{Switching (ours)}:
        For each dataset, one of ``Uncleansed,'' ``Demucs,'' or ``Miipher'' was selected and applied. 
        The selection was based on training data quality, aiming to choose the method that maximized overall quality.
\end{itemize}

Furthermore, to validate the proposed metric, training data quality, we investigated an alternative approach that uses acoustic quality evaluation instead of training data quality.  
In this approach, the preprocessing method that achieves the highest acoustic quality is selected, and the corpus is constructed by selecting data in descending order of acoustic quality.  
The acoustic quality is evaluated using the deep learning model NISQA~\cite{mittag2021nisqa}, where the minimum value among five metrics is used as the evaluation score.

\subsubsection{Evaluation}
We performed the pseudo MOS evaluation according to the procedure described in Section~\ref{sec:3:eval}, aiming to answer the following research questions:
\begin{itemize} \leftskip -3mm \itemsep -0mm
    \item \textbf{Is the ``Switching'' framework effective in improving model performance?}
    \item \textbf{Which criterion is more suitable for determining the data cleansing method: acoustic quality or training data quality?}
\end{itemize}
This evaluation was conducted by calculating  
1) the distribution of the pseudo MOSs for synthetic speech generated by the trained TTS models, and  
2) the number of high-quality speakers, as defined in Section~\ref{sec:3:eval}.

In addition, to evaluate whether the synthetic speech of high-quality speakers truly sounded natural, we conducted an actual MOS test via human evaluation, following the procedure defined in Section~\ref{sec:mos_vs_pseudomos}. This allowed us to verify the effectiveness of the ``Switching'' strategy not only in terms of pseudo MOS but also from a perceptual standpoint.

Finally, we analyzed the proportion of data samples that were processed by each cleansing method in the ``Switching'' scenario. This analysis reveals how each method was selected depending on data characteristics and model robustness, providing further insight into the behavior of the proposed framework.

\subsection{Result}
\newcommand{\swsinglepseudomos}[5]{
    \begin{subfigure}{1.0\linewidth}
    \includegraphics[width=1.0\linewidth]{chapter4/img/#1/switching_pseudo_mos_#2.pdf}
    \vspace{#4}
    \caption{#3}
    \vspace{#5}
    \end{subfigure}
}

\begin{figure}[t]
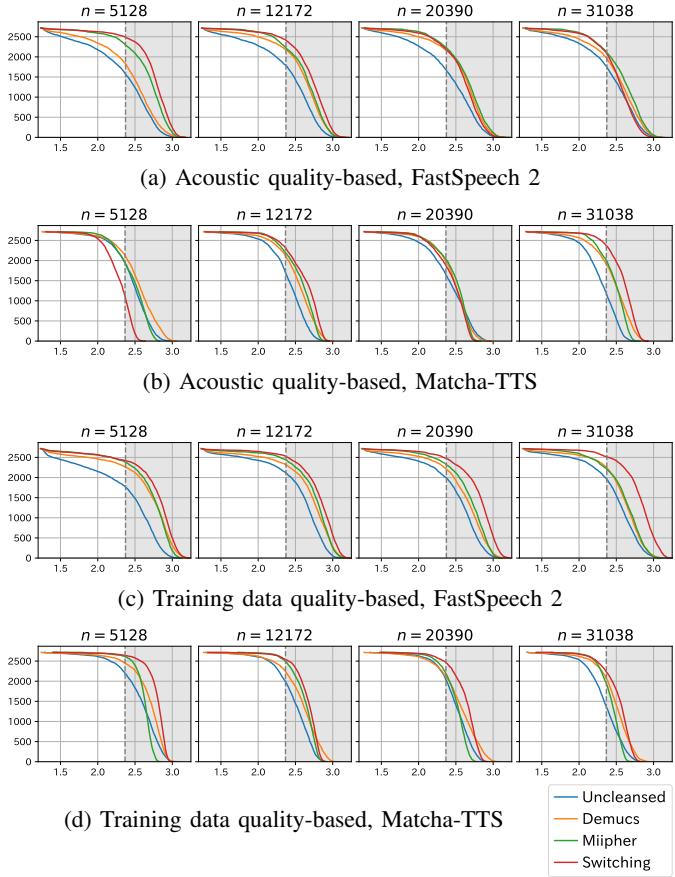

  \centering
  
  \swsinglepseudomos{fs2}{aq}{Acoustic quality-based, FastSpeech~2}{-13pt}{5pt}
  \donehighlight{
  \swsinglepseudomos{matcha}{aq}{Acoustic quality-based, \donemodify{GlowTTS}{Matcha-TTS}}{-13pt}{10pt}
  }
  \swsinglepseudomos{fs2}{ours}{Training data quality-based, FastSpeech~2}{-13pt}{5pt}
  \donehighlight{
  \swsinglepseudomos{matcha}{ours}{Training data quality-based, \donemodify{GlowTTS}{Matcha-TTS}\hspace{40pt}\;}{-45pt}{10pt}
  }
  
  \caption{
    Cumulative histograms of pseudo MOS. 
    Y-axis value indicates number of speakers with higher score than x-axis value.
    The shaded area corresponds to high-quality speakers.
  }
  \label{fig:4:pseudoMOS}
\end{figure}

\begin{table}[t]
  \centering
  \caption{
    Number of overall, seen and unseen speakers. 
    Values of each cell are number of high-quality speakers, all speakers, and ratio of two values, respectively.
  }
  \subfloat[Acoustic Quality based]{
    \begin{tabular}{c|c||c|c}
      $n$ & Method & FastSpeech~2 & \donemodify{GlowTTS}{Matcha-TTS} \\ \midrule \midrule
      \multirow{4}{*}{$5128$}
  &Uncleansed & $1572~(57.8\%)$ & \donemodify{}{$1949~(71.7\%)$}\\
  &Demucs & $1828~(67.2\%)$ & \donemodify{}{$\mathbf{2121~(78.0\%)}$}\\
  &Miipher & $2290~(84.2\%)$ & \donemodify{}{$1928~(70.9\%)$}\\
  &Switching & $\mathbf{2484~(91.4\%)}$ & \donemodify{}{$1104~(40.6\%)$}\\ \midrule
\multirow{4}{*}{$12172$}
  &Uncleansed & $1786~(65.7\%)$ & \donemodify{}{$1698~(62.4\%)$}\\
  &Demucs & $2177~(80.1\%)$ & \donemodify{}{$2114~(77.7\%)$}\\
  &Miipher & $2224~(81.8\%)$ & \donemodify{}{$2196~(80.8\%)$}\\
  &Switching & $\mathbf{2413~(88.7\%)}$ & \donemodify{}{$\mathbf{2315~(85.1\%)}$}\\ \midrule
\multirow{4}{*}{$20390$}
  &Uncleansed & $1661~(61.1\%)$ & \donemodify{}{$1654~(60.8\%)$}\\
  &Demucs & $2155~(79.3\%)$ & \donemodify{}{$1982~(72.9\%)$}\\
  &Miipher & $\mathbf{2223~(81.8\%)}$ & \donemodify{}{$\mathbf{2028~(74.6\%)}$}\\
  &Switching & $2182~(80.3\%)$ & \donemodify{}{$1873~(68.9\%)$}\\ \midrule
\multirow{4}{*}{$31038$}
  &Uncleansed & $1748~(64.3\%)$ & \donemodify{}{$1188~(43.7\%)$}\\
  &Demucs & $1958~(72.0\%)$ & \donemodify{}{$1913~(70.4\%)$}\\
  &Miipher & $\mathbf{2103~(77.3\%)}$ & \donemodify{}{$2007~(73.8\%)$}\\
  &Switching & $2078~(76.4\%)$ & \donemodify{}{$\mathbf{2373~(87.3\%)}$}
    \end{tabular}
    \label{tab:1a}
  }
  \vspace{5pt}
  \subfloat[Training data Quality based]{
    \begin{tabular}{c|c||c|c}
      $n$ & Method & FastSpeech~2 & \donemodify{GlowTTS}{Matcha-TTS} \\ \midrule \midrule
      \multirow{4}{*}{$5128$}
  &Uncleansed & $1764~(64.9\%)$ & \donemodify{}{$2198~(80.8\%)$}\\
  &Demucs & $2263~(83.2\%)$ & \donemodify{}{$2446~(90.0\%)$}\\
  &Miipher & $2379~(87.5\%)$ & \donemodify{}{$2613~(96.1\%)$}\\
  &Switching & $\mathbf{2420~(89.0\%)}$ & \donemodify{}{$\mathbf{2642~(97.2\%)}$}\\ \midrule
\multirow{4}{*}{$12172$}
  &Uncleansed & $2114~(77.7\%)$ & \donemodify{}{$1993~(73.3\%)$}\\
  &Demucs & $2320~(85.3\%)$ & \donemodify{}{$2244~(82.5\%)$}\\
  &Miipher & $2437~(89.6\%)$ & \donemodify{}{$2496~(91.8\%)$}\\
  &Switching & $\mathbf{2524~(92.8\%)}$ & \donemodify{}{$\mathbf{2540~(93.4\%)}$}\\ \midrule
\multirow{4}{*}{$20390$}
  &Uncleansed & $1989~(73.2\%)$ & \donemodify{}{$2035~(74.8\%)$}\\
  &Demucs & $2205~(81.1\%)$ & \donemodify{}{$2101~(77.3\%)$}\\
  &Miipher & $2318~(85.3\%)$ & \donemodify{}{$2193~(80.7\%)$}\\
  &Switching & $\mathbf{2473~(91.0\%)}$ & \donemodify{}{$\mathbf{2464~(90.6\%)}$}\\ \midrule
\multirow{4}{*}{$31038$}
  &Uncleansed & $1955~(71.9\%)$ & \donemodify{}{$1361~(50.1\%)$}\\
  &Demucs & $2241~(82.4\%)$ & \donemodify{}{$2075~(76.3\%)$}\\
  &Miipher & $2204~(81.1\%)$ & \donemodify{}{$1951~(71.8\%)$}\\
  &Switching & $\mathbf{2516~(92.5\%)}$ & \donemodify{}{$\mathbf{2233~(82.1\%)}$}
    \end{tabular}
    \label{tab:1b}
  }
\end{table}

\subsubsection{Pseudo MOS Evaluation}

Fig.~\ref{fig:4:pseudoMOS} shows the cumulative histograms of pseudo MOS for different data cleansing methods and selection metrics.  
When the training data quality is used as the selection metric, the proposed ``Switching'' method consistently outperforms the baseline methods across all corpus sizes and model architectures.  
This consistent superiority demonstrates the effectiveness and robustness of our approach.

In contrast, when acoustic quality is used as the selection metric, the benefit of ``Switching'' becomes less reliable.  
In some conditions, such as $n = 31038$, ``Switching'' even underperforms compared to applying ``Miipher'' uniformly.  
This inconsistency indicates that acoustic quality does not always reflect the actual utility of data for training TTS models.

These observations support our central claim:  
\textit{Training data quality}—defined as the expected contribution of each utterance to downstream model performance—is a more reliable and task-aware criterion for data selection than predetermined criterion such as acoustic quality.  

\begin{figure}[t]
\donehighlight{
\centering
\includegraphics[width=0.9\linewidth]{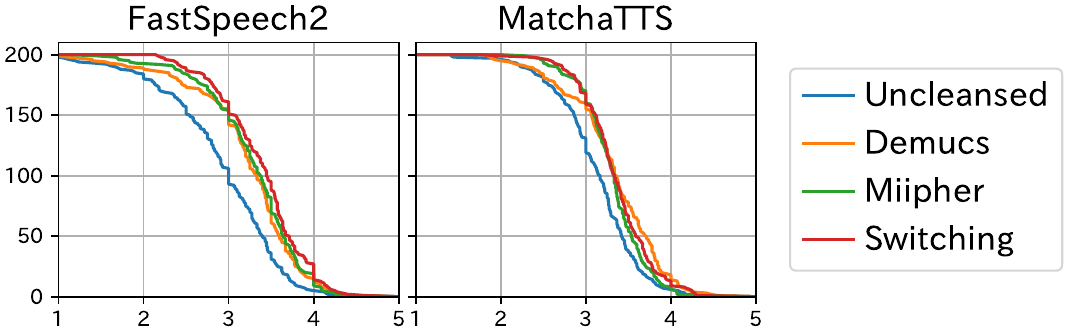}
  \caption{
    Cumulative histograms of actual MOS. 
    Y-axis value indicates number of speakers with higher score than x-axis value.
    The shaded area corresponds to high-quality speakers.
  }
\label{fig:4:subjective}
}
\end{figure}

\subsubsection{Validation on Actual MOS}
We conducted a five-point MOS test following the procedure described in Section~\ref{sec:mos_vs_pseudomos}. In this evaluation, we compared ``Uncleansed,'' ``Demucs,'' ``Miipher,'' and ``Switching,'' all based on the training-data-based selection strategy with a fixed data set size of $n = 12172$.

Fig.~\ref{fig:4:subjective} shows the results.  
For FastSpeech~2, among the individual data cleansing methods, ``Miipher'' performed the best, followed by ``Demucs'' and then ``Uncleansed.''
The proposed method, which dynamically leverages the strengths of each, achieved the best overall performance.
\donemodify{}{In addition, for Matcha-TTS, the ``Switching'' strategy improved upon ``Miipher,'' 
demonstrating the effectiveness of our method across different architectures.}
\donemodify{This result indicates}{These results indicate} that the proposed method is more effective than the conventional methods even in terms of perceptual speech quality.

\donemodify{
On the other hand, for GlowTTS, ``Uncleansed,'' ``Demucs,'' and ``Miipher'' exhibited similar performance in the subjective evaluation, and the proposed method did not yield additional improvements. This suggests that further improvements in preprocessing techniques may be needed to fully realize the benefits of the ``Switching'' approach in models such as GlowTTS.
}{}

\subsubsection{Investigation of the Switching Method} \newcommand{\switchingrate}[2]{
    \begin{subfigure}{1.0\linewidth}
    \includegraphics[width=0.95\linewidth]{chapter4/img/#1/switching_rate.pdf}
    \vspace{0pt}
    \caption{#2}
    \end{subfigure}
}

\begin{figure}[t]
  \centering
  
  \switchingrate{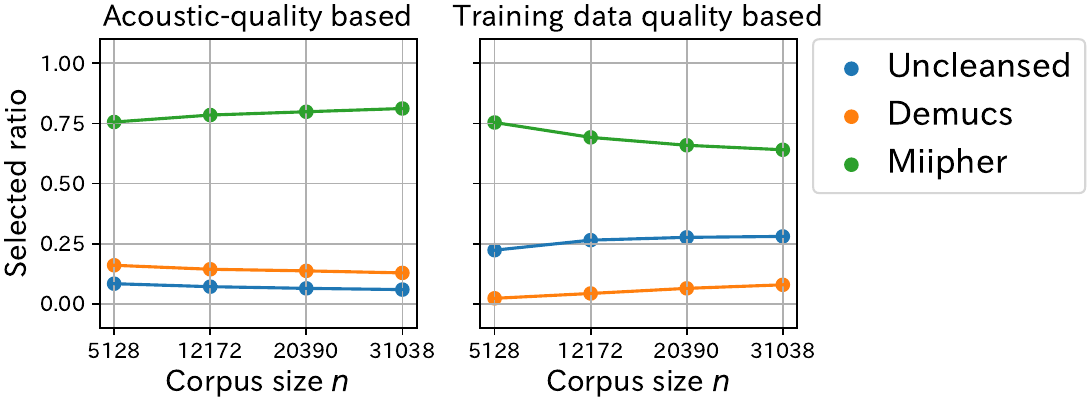}{FastSpeech~2}
  \vfill
  \donehighlight{
  \switchingrate{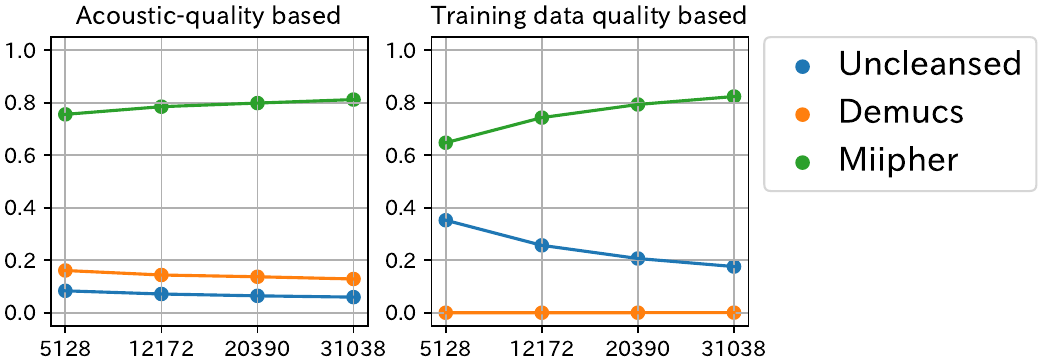}{\donemodify{GlowTTS}{Matcha-TTS}}
  }
  \vspace{-10pt}
  
  \caption{The proportion of each preprocessing method selected in the ``Switching'' approach.}
  \label{fig:4:switching_rate}
\end{figure}

Fig.~\ref{fig:4:switching_rate} presents the selection rates of each preprocessing method used in the "Switching" strategy, across various dataset sizes and model architectures.

Since fewer than $10\%$ of the collected utterances achieved a NISQA score above $4.0$, indicating that the majority of the data were acoustically degraded, ``Miipher'' was selected for most samples due to its strong restoration performance.

However, for \donemodify{FastSpeech 2}{both models and all dataset size $n$}, \donemodify{over $20\%$}{approximately $20\%$–$40\%$} of the utterances were selected without any preprocessing, despite being sourced from noisy YouTube recordings.
This suggests that \donemodify{FastSpeech 2 has}{the models possess} a certain degree of robustness to noise and may benefit from preserving relatively clean segments without risking degradation from unnecessary processing.
\donemodify{}{In other words, leaving some data unprocessed can itself constitute 
an important component of the dynamic selection strategy.}

\donemodify{
In contrast, GlowTTS consistently selected ``Miipher'' across all dataset sizes.
This preference can be attributed to GlowTTS's reliance on internal alignment learning, which makes it more sensitive to noise and alignment errors. Consequently, aggressive restoration becomes necessary to stabilize training, leading to an almost exclusive selection of Miipher.
}{}

These findings highlight that the optimal data cleansing strategy is highly dependent on the architecture and noise tolerance of the target TTS model.
Notably, the selection patterns remained stable across different corpus sizes, underscoring the adaptability and consistency of the ``Switching'' framework in meeting model-specific requirements.

\section{Discussion}
\subsection{Computation Cost}
\donemodify{
One practical concern regarding our framework is the computational cost associated with the training-evaluation loops.  
As described in Section~\ref{sec:chapter3}, our data selection method requires two TTS training cycles: one to estimate training data quality using pseudo MOS scores, and another for final training with the constructed corpus.  
In the case of TTSOps (Section~\ref{sec:chapter4}), the number of training runs increases to $n + 1$, where $n$ is the number of candidate preprocessing methods.  

While this may seem costly at first glance, it reflects the reality of TTS system development.
In practical scenarios, model training is rarely a one-shot process: developers typically go through many rounds of data cleaning, retraining, and hyperparameter tuning to achieve satisfactory results.
TTSOps automates this trial-and-error process by integrating data selection and data cleansing into a closed-loop framework, significantly reducing the need for manual intervention.
As a result, the overall development cycle can become more efficient—even if each run of the framework incurs some additional computation.

As a reference, under our experimental conditions using a single NVIDIA RTX 3090 GPU, each training run of FastSpeech~2 within the TTSOps framework required approximately 2 hours.
Given that the framework involved four training runs in total—corresponding to three candidate preprocessing methods and one final training—the overall training time amounted to approximately 8 hours.
}{
One practical concern regarding our framework is the computational cost associated with the training–evaluation loops.  
As described in Section~\ref{sec:chapter3}, our data selection method requires two TTS training cycles: one to estimate training data quality using pseudo MOS scores, and another for final training with the constructed corpus.  
In the case of TTSOps (Section~\ref{sec:chapter4}), the number of training runs increases to $n+1$, where $n$ is the number of candidate preprocessing methods.  

While this may appear costly at first glance, the actual wall-clock time is reasonably small under modern GPU settings.  
As described in Section~\ref{sec:3:cost}, one training run of FastSpeech~2 on a single NVIDIA H200 GPU required about 1.5 hours, and Matcha-TTS required about 5 hours.  
Even considering multiple runs, the total time remained within a practically feasible range for real-world TTS development.  
Thus, the computational cost of TTSOps is moderate compared to the iterative trial-and-error cycles of data cleaning and retraining that practitioners usually perform manually.  
In practice, system development rarely proceeds in a single shot; rather, developers typically perform many such iterations, and the number of repetitions is often unbounded until satisfactory results are achieved.  
By automating this process, TTSOps makes the development cycle more systematic and predictable, while reducing the need for manual intervention.

Furthermore, we demonstrated in Section~\ref{sec:3:sampling} that a sampling strategy (``Ours-Utt-Rand''), which uses only $10\%$ of the candidate data for the initial training, yields performance comparable to the full-data version.  
This indicates that the computational cost can be further reduced when scaling up to larger corpora, without significantly sacrificing the effectiveness of the proposed method. 
\modify{}{In addition, the sampling strategy also suggests that a small-scale validation experiment can serve as a practical way to preliminarily assess the effectiveness of TTSOps for a given model, 
which can be beneficial in real-world deployment scenarios.}
\modify{In addition}{Moreover}, the $n$ training runs required for evaluating 
different preprocessing methods in TTSOps are mutually independent 
and can therefore be executed in parallel. 
Thus, when multiple GPUs are available, the total time can be 
further reduced almost proportionally to the number of GPUs.
}

\subsection{Scalability and Language-Agnostic Design}
A key strength of the proposed framework lies in its fully automated and language-agnostic design.  
Because our method does not rely on language-specific heuristics or human intervention, it can be easily scaled to larger corpora and adapted to various languages.  
In our experiments, we successfully applied the method to a large-scale Japanese dataset, demonstrating its practicality under real-world conditions.  
\donemodify{
In particular, the MOS prediction model used in our method was trained on English and Chinese data.  
Achieving promising results on Japanese---an out-of-domain language for the model---serves as experimental evidence for the language-independence of our approach.  
Further validation across different languages and datasets remains an important direction for future work.
}{
Obtaining promising results in a language not included in the training data of the MOS prediction model provides experimental evidence for the language-independence of the approach.  
Furthermore, the experiments on English presented in Section~\ref{sec:3:language} also yielded promising results, further emphasizing the language-agnostic nature of the proposed method.
}
\modify{}{
Future work is expected to extend the evaluation to additional languages to further confirm the language-agnostic effectiveness of the method.
}

\donemodify{
\subsection{Limitations of Pseudo MOS and Future Prospects}
\label{sec:mos_lim}
While the correlation between pseudo MOS and actual MOS was limited for GlowTTS, this issue is expected to diminish as MOS prediction models continue to advance.  
Indeed, international competitions such as the VoiceMOS Challenge are held annually~\cite{huang2022voicemos, cooper2023voicemos, huang2024voicemos}, promoting rapid progress in prediction accuracy.  
These developments have enabled more fine-grained evaluations, such as distinguishing subtle differences in high-MOS regions and conducting zero-shot assessments.

}{
\subsection{Prospects of Pseudo MOS and Future Extensions}
\label{sec:mos_lim}
As discussed in Section~\ref{sec:mos_vs_pseudomos}, both FastSpeech~2 and Matcha-TTS 
exhibited strong correlations between pseudo MOS and actual MOS, 
although further improvements in performance can be expected in the future.  
International competitions such as the VoiceMOS Challenge~\cite{huang2022voicemos, cooper2023voicemos, huang2024voicemos} 
have been driving rapid advances in prediction accuracy, 
enabling more fine-grained evaluations such as distinguishing subtle differences in high-MOS regions 
and conducting zero-shot assessments.  
These developments are expected to further enhance the effectiveness of our proposed method.
}

A notable feature of our TTSOps framework is its flexibility: the evaluation-in-the-loop mechanism is not restricted to pseudo MOS.  
It can be extended to incorporate alternative evaluation metrics (e.g., prosody, emotion), thereby broadening its applicability according to the specific training objectives.

\subsection{Importance of Switching Data Cleansing Method}
Although ``Miipher'' was selected for the majority of utterances in our experiments, this does not suggest that applying Miipher uniformly constitutes an optimal strategy.  
Indeed, the documentation for LibriTTS-R~\cite{koizumi2023libritts} reports that approximately $5\%$ of samples failed to be successfully restored using Miipher\footnote{\url{https://www.openslr.org/141/}}.  
Our switching strategy enables the automatic exclusion of such degraded samples and the selection of more suitable alternatives on a per-utterance basis.

As new enhancement models continue to emerge—targeting specific issues such as background noise, codec artifacts, or reverberation—the importance of dynamic, utterance-specific preprocessing is expected to increase.  
Our framework is designed to accommodate this evolving landscape, maintaining robustness in the face of both the strengths and limitations of future enhancement techniques.

From this perspective, the proposed ``Switching'' mechanism represents a forward-compatible design that anticipates the growing diversity and complexity of speech restoration models.  
It offers a principled and scalable solution for optimizing data cleansing strategies automatically, without reliance on static or manually crafted pipelines.

\section{Conclusion}
\label{sec:conclusion}
This paper presented \textbf{TTSOps}, a fully automated, closed-loop corpus optimization framework for training multi-speaker text-to-speech (TTS) models from large-scale, noisy, and uncurated web data. Unlike conventional pipelines that rely on static data cleansing and model-agnostic selection criteria, TTSOps unifies evaluation-in-the-loop data selection and utterance-wise data cleansing switching, both guided by a novel criterion: the training data quality—defined as the estimated contribution of each utterance to downstream TTS performance.

By integrating model-aware evaluation and adaptive data cleansing into the training loop, TTSOps effectively constructs high-quality, speaker-diverse corpora from real-world ``dark data'' such as YouTube videos. Extensive experiments using two TTS architectures demonstrated that TTSOps consistently outperforms acoustic-quality-based baselines in terms of naturalness, speaker coverage, and speaker diversity. Furthermore, the effectiveness of the framework was validated not only through pseudo MOS but also by human-rated MOS evaluations.

Importantly, the framework adapts to the robustness characteristics of the target TTS model and flexibly selects the most effective data cleansing method for each utterance, enabling principled and scalable corpus construction. \donemodify{While modest discrepancies between pseudo and actual MOS were observed for certain models (e.g., GlowTTS), the increasing accuracy of automatic speech quality predictors promises further gains.}{
The computational cost was shown to remain moderate under realistic GPU settings and can be further reduced by applying sampling strategies. 
Moreover, experiments on English demonstrated that the method generalizes beyond Japanese, supporting its language-agnostic design.}

In summary, TTSOps offers a practical and forward-compatible approach to data-centric TTS development, bridging the gap between uncurated web data and high-quality synthetic speech. Future work includes scaling the framework to multilingual and cross-domain settings, incorporating alternative quality metrics (e.g., prosody or expressiveness), and exploring lightweight evaluation strategies to further reduce computational cost.

\bibliographystyle{IEEEtran}
\bibliography{reference}

\begin{IEEEbiography}[{\includegraphics[width=1in,height=1.25in,clip,keepaspectratio]{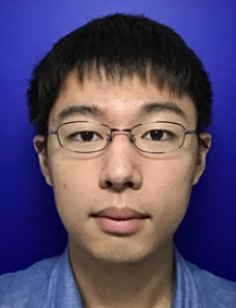}}]{Kentaro Seki} received the B.E. and the M.E. degree in 2022 and 2024, respectively from the University of Tokyo, Tokyo, Japan, where he is currently working toward the Ph.D. degree.
His research interests include speech synthesis, speech enhancement, and audio signal processing.
He is a student member of several organizations including IEEE, IEEE Signal Processing Society (SPS), and Acoustical Society of Japan (ASJ).
He was the recipient of several awards and grants, including IEEE SPS Travel Grant for IEEE ICASSP 2023, Google Travel Grants for Students in East Asia, the best student presentation award of ASJ in 2022.
He was also the recipient of the Research Fellowship for Young Scientists (DC1) from the Japan Society for the Promotion of Science (JSPS) in 2024.
\end{IEEEbiography}

\begin{IEEEbiography}[{\includegraphics[width=1in,height=1.25in,clip,keepaspectratio]{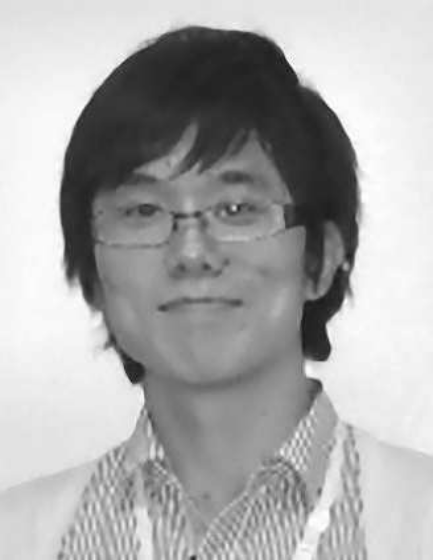}}]{Shinnosuke Takamichi} received the Ph.D. degree from the Graduate School of Information Science, Nara Institute of Science and Technology, Japan, in 2016. He is currently an Associate Professor at Keio University, Japan. He has received more than 20 paper/achievement awards including the 2020 IEEE Signal Processing Society Young Author Best Paper Award.
\end{IEEEbiography}

\begin{IEEEbiography}[{\includegraphics[width=1in,height=1.25in,clip,keepaspectratio]{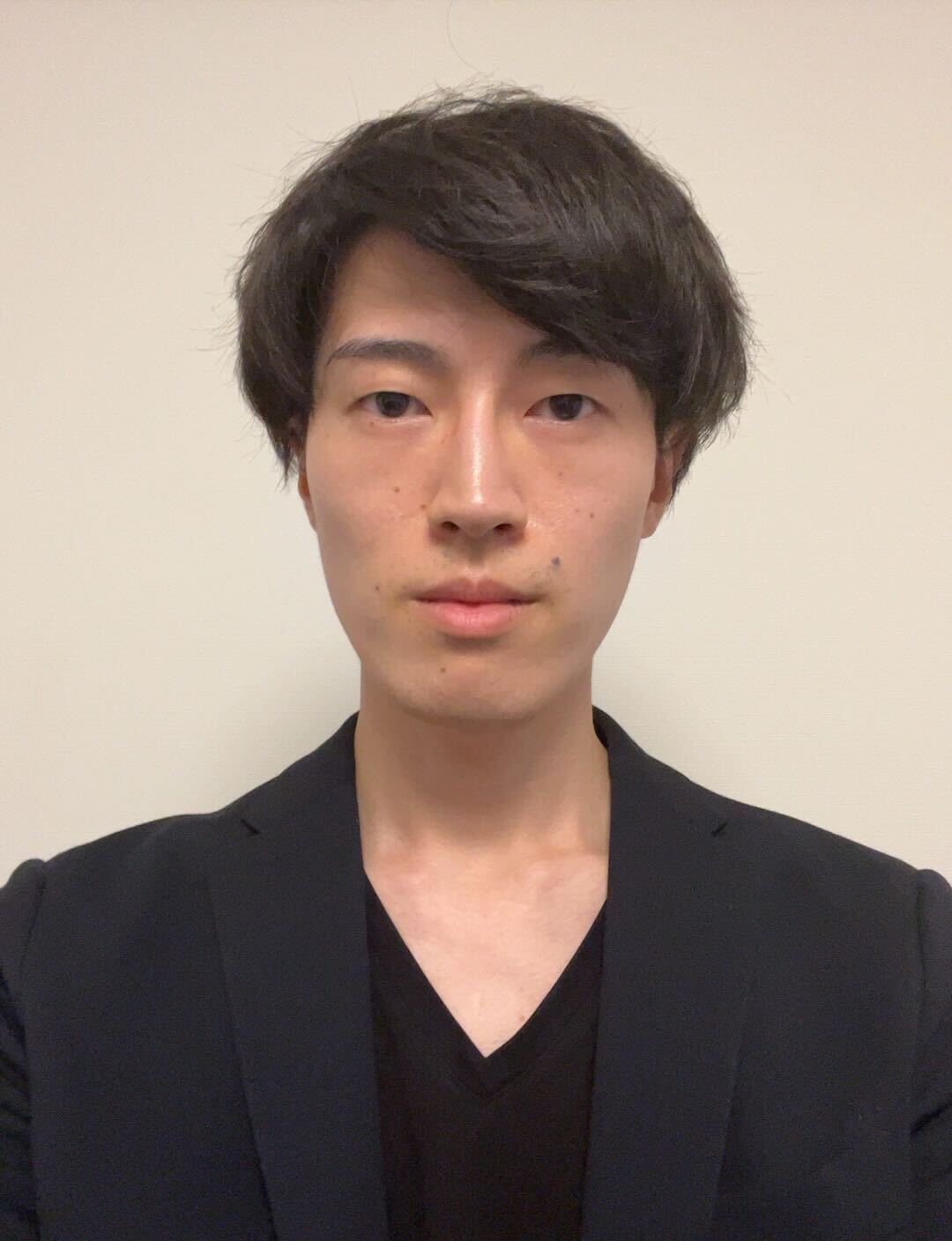}}]{Takaaki Saeki}
received the Ph.D. degree from the University of Tokyo, Japan, in 2024.
His research interests include speech synthesis, speech representation learning, and machine learning. He is currently a Research Scientist at Google DeepMind, USA, where he works on multimodal language modeling. He was the recipient of several awards, including Yamashita SIG Research Award from the Information Processing Society of Japan, the Best Paper Award from the Institute of Electronics, Information and Communication Engineers (IEICE), Japan.
\end{IEEEbiography}

\begin{IEEEbiography}[{\includegraphics[width=1in,height=1.25in,clip,keepaspectratio]{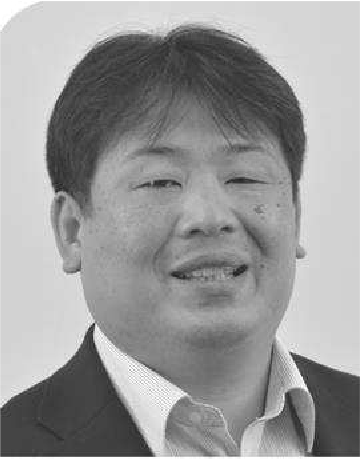}}]{Hiroshi Saruwatari} received the B.E., M.E., and Ph.D. degrees from Nagoya University, Nagoya, Japan, in 1991, 1993, and 2000, respectively. In 1993, he joined SECOM IS Laboratory, Tokyo, Japan, and in 2000, Nara Institute of Science and Technology, Ikoma, Japan. Since 2014, he has been a Professor with The University of Tokyo, Tokyo, Japan. His research interests include statistical audio signal processing, blind source separation, and speech enhancement. He has put his research into the world's first commercially available independent-component-analysis-based BSS microphone in 2007. He was the recipient of several paper awards from IEICE in 2001 and 2006, from TAF in 2004, 2009, 2012, and 2018, from IEEE-IROS2005 in 2006, and from APSIPA in 2013 and 2018, and also the DOCOMO Mobile Science Award in 2011, Ichimura Award in 2013, Commendation for Science and Technology by the Minister of Education in 2015, Achievement Award from IEICE in 2017, and Hoko-Award in 2018. He has been professionally involved in various volunteer works for IEEE, EURASIP, IEICE, and ASJ. Since 2018, he has been an APSIPA Distinguished Lecturer.
\end{IEEEbiography}

\vfill

\end{document}